\documentclass[reprint,superscriptaddress,nobibnotes,longbibliography,aps]{revtex4-2}

\usepackage{dcolumn}
\usepackage{amssymb, amsmath, esint}
\usepackage{physics}
\usepackage{bm}
\usepackage{graphicx}
\usepackage[colorlinks,allcolors=blue]{hyperref}
\usepackage{url}
\usepackage{verbatim}

\begin{document}

\title{Cavity-QED-controlled two-dimensional Moiré Excitons without twisting}

\author{Francesco Troisi}
\email{francesco.troisi@mpsd.mpg.de}
\affiliation{Max Planck Institute for the Structure and Dynamics of Matter and Center for Free-Electron Laser Science, Luruper Chaussee 149, 22761, Hamburg, Germany}

\author{Hannes Hübener}
\email{hannes.huebener@mpsd.mpg.de}
\affiliation{Max Planck Institute for the Structure and Dynamics of Matter and Center for Free-Electron Laser Science, Luruper Chaussee 149, 22761, Hamburg, Germany}

\author{Angel Rubio}
\email{angel.rubio@mpsd.mpg.de}
\affiliation{Max Planck Institute for the Structure and Dynamics of Matter and Center for Free-Electron Laser Science, Luruper Chaussee 149, 22761, Hamburg, Germany}
\affiliation{Initiative for Computational Catalysis (ICC), The Flatiron Institute, 162 Fifth Avenue, New York, NY 10010, United States}

\author{Simone Latini}
\email{simola@dtu.dk}
\affiliation{Department of Physics, Technical University of Denmark, 2800 Kgs. Lyngby, Denmark}
\affiliation{Max Planck Institute for the Structure and Dynamics of Matter and Center for Free-Electron Laser Science, Luruper Chaussee 149, 22761, Hamburg, Germany}

\begin{abstract}
We propose an all-optical Moiré-like exciton confinement by means of spatially periodic optical cavities. 
Such periodic photonic structures can control the material properties by coupling the matter excitations to the confined photons and their quantum fluctuations. We develop a low energy non-perturbative quantum electro-dynamical description of strongly coupled excitons and photons at finite momentum transfer.
We find that in the classical limit of a laser driven cavity the induced optical confinement directly emulates Moiré physics.
In a dark cavity instead, the sole presence of quantum fluctuations of light generates a sizable renormalization of the excitonic bands and effective mass. We attribute these effects to long-range cavity-mediated exciton-exciton interactions which can only be captured in a non-perturbative treatment. With these findings we propose spatially structured cavities as a promising avenue for cavity material engineering.
\end{abstract}

\pacs{}
\maketitle

%%%%%%%%%%%%%%%%%%%%%%%%%%%%%%%%%%%%%%%%%%%%%%%%%
% Introduction
%%%%%%%%%%%%%%%%%%%%%%%%%%%%%%%%%%%%%%%%%%%%%%%%%
\begin{figure*}[t!]
    \centering
    \includegraphics[width=\linewidth]{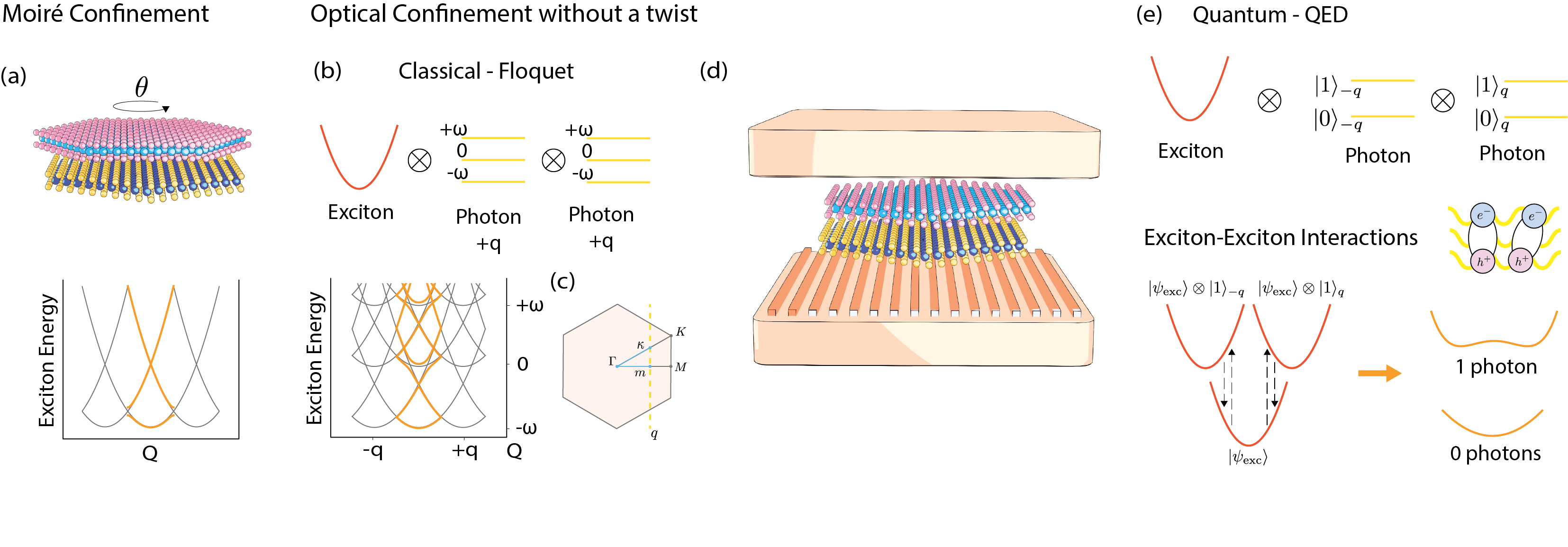}
    \caption{Schematics of the Moiré band formation and cavity-mediated interaction in a TMD bi-layer hetero-structure. (\textbf{a}) A structural twisted bi-layer TMD hetero-structure forms a Moiré super-lattice, leading to spatial modulation of the excitonic states. (\textbf{b}) Untwisted hetero-structure embedded into a classically-driven spatially structured cavity, the light-matter interaction generates an all-optical Moiré potential. (\textbf{c}) Representation of the \textit{k}-points path used in the spectral function. We use the standard $M-\Gamma-K$ path, which we shorten to $m-\Gamma-\kappa$, where $m, \kappa$ are taken at the value of the photon momentum along the aforementioned path. (\textbf{d}) Representation of an untwisted TMD hetero-structure embedded in a spatially structured optical cavity. (\textbf{e}) When the hetero-structure is untwisted (no standard Moiré potential) and embedded into a dark spatially structured cavity (where the vacuum fluctuation play a role), light–matter coupling mediates exciton-exciton interactions and re-normalizes the excitonic mass.}
    \label{fig:cartoon}
\end{figure*}

\section{Introduction}\label{sec:introduction}
Two-dimensional Van der Waals materials have emerged as a versatile platform for opto-electronic devices thanks to their highly tuneable properties, such as optical and electronic band gaps and dielectric response~\cite{VdW, Mak2016, Novoselov2016, Shanmugam2022, kennes2021moire}.
Among this class of materials, transition metal dichalcogenides (TMDs) have attracted particular interest due to their strongly bound excitons and related optical activity in the visible range~\cite{mak2010atomically, latini2015excitons, gjerding2021recent}.
Excitons do indeed dominate the dynamics of several phenomena of TMDs, such as valley polarization and non-linear optical response~\cite{RevModPhys.90.021001, ren20252d}. Hence, controlling the behavior of excitons offers a direct route to program optical functionalities in this class of materials.
Twist engineering~\cite{bistritzer2011moire, cao2018unconventional, Hennighausen_2021}, a technique in which atomically thin TMD layers are stacked with a predefined twist angle, allows for spatial control of excitons and consequent modification of their optical activity~\cite{tunable_phases}. This is possible because twisting two-dimensional (2D) crystals with respect to each other leads to the formation of Moiré patterns, which generate spatially periodic electrostatic potentials strong enough to influence the motion and confinement of the otherwise free excitons, c.f. Fig.~\ref{fig:cartoon}(a).
Aside from excitonic control, twist engineering, also dubbed twistronics, allows for generating novel quantum phases~\cite{2d_coulomb, kennes2021moire}, including correlated insulator states~\cite{Xiong2023}, unconventional superconductivity~\cite{Cao2018} and fractional Chern insulators~\cite{Sharpe2019}.

In this work we propose an alternative strategy for spatial confinement of excitons which utilizes optical cavities instead of twist engineering. Optical cavities confine the electromagnetic field in a small volume, strengthening its intensity and making it possible to strongly couple to excitons of embedded TMDs.
Such interaction can in turn lead to the formation of exciton-polaritons~\cite{cavity_control, Flatten2016, Zhang2020, Svendsen2021, Datta2022Full}.
Previous works show that due to the strong light-matter interaction, planar optical cavities allow to control the creation and mixing of polaritons made of composite excitons~\cite{cavity_control}, as well as excitons, phonon and photon quasi-particles, the so-called phonoriton~\cite{Phonoritons}. 
Cavity confinement not only alters the excitonic spectrum~\cite{Microcavity_Exciton_Polaritons, Kasprzak2006} but can also give rise to new equilibrium quantum phases due to the coupling of the host material with the vacuum fluctuations of light~\cite{Appugliese2022, Sentef2018, Mazza2019, Hubener2024, ferroelectric_gs, light_fluid, VinasBostrom2023, Lu2025}.
Here we propose using spatially structured optical cavities, such as those sketched in Fig.~\ref{fig:cartoon}, to produce an all-optical Moiré-like exciton confinement without twisting.
This is a different paradigm compared to previous works that deal with exciton-polaritons in grated systems as they aim to reshape the properties of emitted light~\cite{Zhang2020, Sarkar2024}. With our approach, instead, we demonstrate that we can control the matter properties by specifically structuring the excitonic quasi-particle, using both quantum fluctuations and classical fields.

We analyze the role of cavity-mediated interactions in a prototypical type-II MoSe$_2$/WSe$_2$ hetero-structure embedded in a planar cavity setup. 
We consider both an unstructured planar cavity and a structured (grated) cavity. The former can be described as a single effective mode of the electromagnetic field, where light carries no momentum, while the latter requires a multi-mode description and allows for momentum transfer between light and matter.
Our theoretical framework builds on the one hand on the methodology for the first-principles treatment of excitons in Moiré potentials established in Ref.~\cite{tunable_phases} and on the other hand on a low energy quantum electro-dynamical (QED) Hamiltonian approach for the coupling of excitons to the cavity~\cite{cavity_control, qed_hamiltonian}.
More specifically, we solve the Mott-Wannier equation in momentum space, to obtain the low-energy excitonic states.
Subsequently we derive a low energy QED Hamiltonian which can describe both the electrostatic Moiré potential arising from twisting the bi-layer and the coupling to the photonic modes by using an excitonic representation of the many-body QED problem.
Finally, we perform the full diagonalization of the QED Hamiltonian to obtain the hybrid exciton-polariton states, from which we predict the excitonic dispersion, via the excitonic spectral function, and the optical properties of the cavity-matter system by computing the interacting optical linear absorption spectra.

We find that spatially unstructured cavities, where the electromagnetic field carries no momentum, can alter the twist induced excitonic confinement from the Moiré potential when the cavity-mode is resonant with the excitonic transition.
Conversely, spatially structured cavities, where momentum can be exchanged between light and matter, can induce optical confinement and emulate the Moiré potential when driven with a classical electromagnetic field (i.e. a laser), even in the \textit{absence} of twist in the embedded bi-layer.
For dark spatially structured cavities, instead, where the light-matter coupling arises from the quantum fluctuations of the electromagnetic field inside the cavity, we find the emergence of cavity-induced exciton-exciton interactions in untwisted bi-layers. This leads to both excitonic confinement and mass renormalization and consequently to a modification of the optical properties of the material in equilibrium, going beyond what is possible to achieve with twisting.

%The paper is organized as follows: Section \ref{sec:results} describes the results of the paper, starting from the formulation of the QED Hamiltonian (Sec.~\ref{subsec:theory}) and then presenting the numerical calculations for both a classical (Sec.~\ref{subsec:classical_cavity}) and quantum (Sec.~\ref{subsec:dark_cavity}) treatment of the radiation field. Finally, section \ref{sec:methods} describes structured and unstrucutred cavities as well as relevant methods and approximations used in the paper.

%%%%%%%%%%%%%%%%%%%%%%%%%%%%%%%%%%%%%%%%%%%%%%%%%
% Results
%%%%%%%%%%%%%%%%%%%%%%%%%%%%%%%%%%%%%%%%%%%%%%%%%
\section{Results}\label{sec:results}

\subsection{Theory} \label{subsec:theory}
To study the behavior of Moiré excitons in an optical cavity, we first model the two uncoupled systems (excitons and photons) and subsequently describe their interaction. Then, we formulate and discuss the QED Hamiltonian for Moiré excitons, which constitutes one of the main results of the paper.
All throughout this work we use atomic units.

The uncoupled matter Hamiltonian $\hat{H}_M$ of a twisted bi-layer hetero-structure can be formulated in an excitonic representation following Ref.~\cite{tunable_phases} as
\begin{equation}
    \label{eq:h_matter_x_op}
    \begin{aligned}
        & \hat{H}_M = \sum_{ll', \nu}\sum_{i \in C, j \in V} \sum_{\bm{Q}} \mathcal{E}_{ll',ij,\bm{Q}}^{\nu} \hat{X}_{ll',ij,\bm{Q}}^{\nu \dagger} \hat{X}_{ll',ij,\bm{Q}}^{\nu} + \\
        &   \sum_{ll', \nu} \sum_{i \in C, j \in V} \sum_{\bm{Q}, \bm{q}} \mathcal{M}_{ll',ij,\bm{q}}^{\nu} \hat{X}_{ll',ij,\bm{Q+q}}^{\nu \dagger} \hat{X}_{ll',ij,\bm{Q}}^{\nu}
    \end{aligned}
\end{equation}
where $i,j$ are band indexes which span over conduction ($C$) or valence ($V$) band states, and $l,l'$ are layer indexes. The operators $\hat{X}^\dagger, \hat{X}$ create (annihilate) an exciton between any pair of bands of either the same layer (when $l = l'$, in which case one has intra-layer excitons) or different layers (inter-layer excitons), $\bm{Q}$ is the momentum associated with the center of mass of the exciton and $\bm{q}$ is the momentum transferred by the Moiré potential. The index $\nu$ refers to the bound state (i.e. $1s, 2s ...$).
$\mathcal{E}_{ll',ij,\bm{Q}}^{\nu} = \frac{\bm{Q}^2}{2m_{ll',ij}} + E_{g,ll'} + E_{b,ll'}^{\nu}$ encodes the dispersion relation of a free exciton, which we assume parabolic, where $m_{ll',ij}$ is the excitonic effective mass, $E_{g,ll'}$ is the energy gap and $E_{b,ll'}^{\nu}$ is the binding energy. $\mathcal{M}$ is the matrix element of the Moiré potential in the excitonic basis, describing the Moiré scattering of excitons with different momenta. Refer to the Appendix~\ref{app:qed_hamiltonian} for a detailed derivation.

We describe the uncoupled light system with a Hamiltonian consisting of a set of effective harmonic oscillators (the photon modes of the cavity):
\begin{equation}
    \label{eq:h_ph}
    \hat{H}_{ph} = \sum_{\bm{\bar{q}},\lambda} \omega_{\bm{\bar{q}}} \left(\hat{a}_{\bm{\bar{q}}, \lambda}^\dagger \hat{a}_{\bm{\bar{q}}, \lambda} + \frac{1}{2}\right)
\end{equation}
where $\omega_{\bm{\bar{q}}}$ represents the energy of the photon mode $\bm{\bar{q}}$ and $\lambda$ is the polarization.
$\bm{\bar{q}}$ is the momentum of the photon in the $xy$ plane of the 2D material. In a planar setup, electromagnetic waves propagate in the in-plane direction while they are standing waves in the $z$ direction. We assume that the matter couples with the in-plane component of the electric field and therefore only consider modes for which it is finite. The momentum of the photon in the $z$-direction is set by the standing wave condition for the fundamental cavity mode. Additionally, when a periodic grating is present, the $xy$ plane momentum is finite and determined by the periodicity of the grating.
When studying optical properties of materials in far field, one usually makes the approximation that the light does not carry any momentum, the long-wavelength approximation (LWA). We recently showed that also within a QED framework the LWA is applicable~\cite{qed_hamiltonian} and reduces the description of the electromagnetic Hamiltonian to one with a single effective  mode at $\bar{\bf{q}}=0$. In practice the LWA implies that an electron cannot scatter to another $k$-point following the absorption of a photon (vertical transition in $k-$space).
In contrast, if the cavity is spatially structured with a grating, the photon modes can acquire a finite momentum $\bm{\bar{q}}$ and couple matter excitations with different momenta. As shown later, this can give rise to a non-local interaction of excitons in $k$-space. Note that even though the construction of effective cavity modes provided in Ref.~\cite{qed_hamiltonian} is within the LWA, we here apply a similar procedure to construct the few effective collective modes in Eq.~\ref{eq:h_ph} for which the in-plane momenta are not averaged around $\bar{\bf{q}}=0$, but around the finite momenta $\bm{\Bar{q}}$ set by the cavity.
We stress here that the term "effective", when referring to the cavity modes, indicates that such modes represent a summation over a continuum of modes centered around a certain momentum $\bm{\Bar{q}}$. Working with effective modes is necessary to guarantee a finite light-matter coupling strength in the bulk limit of extended cavity-matter systems when working with a finite number of modes for the description of the electromagnetic field.

To describe the light-matter coupling, we start by performing the canonical momentum substitution $\bm{\hat{p}} \rightarrow \bm{\hat{p} + \hat{A}}$ to the uncoupled matter Hamiltonian, obtaining the second-quantized Pauli-Fierz Hamiltonian~\cite{qed_hamiltonian}.
The Hamiltonian in the excitonic representation reads (c.f. the Appendix~\ref{app:qed_hamiltonian} for the complete derivation):
\begin{widetext}
    \begin{equation}
        \label{eq:h_qed_final}
        \begin{aligned}
            & \hat{H} = \sum_{\bm{\bar{q}},\lambda} \omega_{\bm{\bar{q}}} \left(\hat{a}_{\bm{\bar{q}}, \lambda}^\dagger \hat{a}_{\bm{\bar{q}}, \lambda} + \frac{1}{2}\right) + \sum_{ll', \nu} \sum_{i \in C, j \in V} \sum_{\bm{Q}} \left( \mathcal{E}_{ll',ij,\bm{Q}}^{\nu} \hat{X}_{ll',ij,\bm{Q}}^{\nu \dagger} \hat{X}_{ll',ij,\bm{Q}}^{\nu} + \sum_{\bm{q}} \mathcal{M}_{ll',ij,\bm{Q q}}^{\nu} \hat{X}_{ll',ij,\bm{Q+q}}^{\nu \dagger} \hat{X}_{ll',ij,\bm{Q}}^{\nu} \right) + \\
            & \sum_{\bm{\bar{q}}, \lambda} \tilde{A}_{0, \bm{\bar{q}}} \sum_{ll', \nu} \sum_{i \in C, j \in V} \sum_{\bm{Q}} \Biggl[ \mathcal{B}_{ll',ij,\bm{Q \bar{q}}}^{\nu \lambda} \hat{X}_{ll',ij,\bm{Q+\bar{q}}}^{\nu \dagger} \hat{X}_{ll',ij,\bm{Q}}^{\nu} \left(\hat{a}_{\bm{\bar{q}},\lambda}^\dagger + \hat{a}_{-\bm{\bar{q}}, \lambda}\right) + \mathcal{I}_{ll',ij,\bm{Q \bar{q}}}^{\nu, \lambda} \hat{X}_{ll',ij,\bm{\bar{q}}}^{\nu \dagger} \left(\hat{a}_{\bm{\bar{q}},\lambda}^\dagger + \hat{a}_{-\bm{\bar{q}}, \lambda}\right) + h.c. \Biggr] \\ 
            %&  \sum_{\lambda \lambda'} \sum_{\bm{\Bar{q}},\bm{\Bar{q}}'} \frac{\Tilde{A}_{0, \bm{\Bar{q}}} \Tilde{A}_{0, \bm{\Bar{q}}'}}{2} \sum_{ll'} \sum_{i \in C, j \in V} \sum_{\bm{Q}, \nu} \mathcal{D}_{ll';ij;\bm{Q \Bar{q} \Bar{q}'}}^{\nu; \lambda \lambda'} \hat{X}_{ll';ij;\bm{Q - \Bar{q} + \Bar{q}'}}^{\nu \dagger} \hat{X}_{ll';ij;\bm{Q}}^{\nu} \left(\hat{a}_{\bm{\Bar{q}}; \lambda}^\dagger + \hat{a}_{-\bm{\Bar{q}}; \lambda}\right) \left(\hat{a}_{\bm{\Bar{q}}'; \lambda'}^\dagger + \hat{a}_{-\bm{\Bar{q}}'; \lambda'}\right) + \\
            %& \sum_{\lambda \lambda'} \sum_{\bm{\Bar{q}},\bm{\Bar{q}}'} \frac{\Tilde{A}_{0, \bm{\Bar{q}}} \Tilde{A}_{0, \bm{\Bar{q}}'}}{2} \sum_{ll'} \sum_{i \in C, j \in V} \sum_{\bm{k}, \nu} \mathcal{S}_{ll';ij;\bm{\Bar{q},\Bar{q}'}}^{\nu, \lambda \lambda'} \hat{X}_{ll';ij;\bm{\Bar{q}-\Bar{q}'}}^{\nu \dagger} \left(\hat{a}_{\bm{\Bar{q}};\lambda}^\dagger + \hat{a}_{-\bm{\Bar{q}}; \lambda}\right) \left(\hat{a}_{\bm{\Bar{q}}'; \lambda'}^\dagger + \hat{a}_{-\bm{\Bar{q}}'; \lambda'}\right) + h.c.
        \end{aligned}
    \end{equation}
\end{widetext}

\noindent where $\tilde{A}_{0,\bm{\bar{q}}} = \frac{A_{0, \bm{\bar{q}}}}{\sqrt{V_{\textit{eff}, \bm{\bar{q}}}}}$, $A_{0, \bm{\bar{q}}}$ is the coupling strength of the mode $\bm{\bar{q}}$ and $V_{\textit{eff}, \bm{\bar{q}}}$ is the effective mode volume. The first line of the Hamiltonian contains the uncoupled photon and the uncoupled matter, while the second represents the paramagnetic bi-linear coupling between photon modes and excitons.
Note that we absorbed the diamagnetic term into the uncoupled photon Hamiltonian by performing a Bogoliubov transformation~\cite{rokaj2022free, CavityGraphene}.
$\mathcal{B}$ and $\mathcal{I}$ are the matrix elements describing the coupling to the matter momenta~\cite{qed_hamiltonian} in momentum-conserving exciton-photon interactions and are defined in Appendix~\ref{subsec:qed_c}. The former allows an exciton to scatter to another \textit{k}-point after absorbing or emitting a photon, the implications of which will be discussed in depth in the next sections. Importantly, $\mathcal{B} = 0$ when $\bm{\bar{q}} = \bm{0}$ (see Appendix~\ref{subsec:transition_me}).
The matrix elements $\mathcal{I}$ couple the material ground state to the light by creating an exciton. 
Hence, while the term $\mathcal{B}$ explicitly conserves the number of excitons, $\mathcal{I}$ deals with the creation or destruction of such particles.
In order to access the polaritonic states via full diagonalization, the Hamiltonian in Eq.~\ref{eq:h_qed_final} is projected onto a combined light-matter product state, with excitonic states and the many-body ground state for the matter~\cite{cavity_control}, which are written in a Slater determinant representation, and number states for each photonic mode.
This basis keeps the N-particles electronic states explicit, so that even if the number of excitons is not conserved, the total number of particles is fixed.

In Eq.~\ref{eq:h_qed_final} we observe that the standard Moiré potential $\mathcal{M}$ and the first term of the bi-linear coupling $\mathcal{B}$ share the same excitonic operators $\hat{X}^\dagger, \hat{X}$, in case $\bm{\bar{q}}=\bm{q}$.
For the Moiré potential, $\bm{q}$ refers to the super-lattice Moiré periodicity. For the bi-linear coupling $\bm{\bar{q}}$ refers to the periodicity of the electromagnetic field inside the cavity.
Despite having different physical meanings, both indices determine a lattice periodicity, which modifies the symmetry experienced by the excitons.
To highlight this equivalence, we let $\bm{\bar{q}} = \bm{q}$ and rewrite the interaction terms of the Hamiltonian in Eq.~\ref{eq:h_qed_final} (which also includes the non-conserving bilinear coupling $\mathcal{I}$) as:
\begin{equation}
    \label{eq:h_interaction}
    \begin{aligned}
        & \hat{H}_{\textit{int}} = \sum_{ll', \nu} \sum_{i \in C, j \in V} \sum_{\bm{Q}, \bm{q}} \Biggr[ \hat{X}_{ll',ij,\bm{Q+q}}^{\nu \dagger} \hat{X}_{ll',ij,\bm{Q}}^{\nu} \\
        & \left. \left(\mathcal{M}_{ll',ij,\bm{Q,q}}^{\nu} + \tilde{A}_{0,\bm{q}} \sum_{\lambda}  \mathcal{B}_{ll',ij,\bm{Q q}}^{\nu \lambda} \left(\hat{a}_{\bm{q}, \lambda}^\dagger + \hat{a}_{-\bm{q}, \lambda}\right) \right) \right. + \\
        & \tilde{A}_{0,\bm{q}} \sum_{\lambda} \mathcal{I}_{ll',ij,\bm{Q q}}^{\nu, \lambda} \hat{X}_{ll',ij,\bm{q}}^{\nu \dagger} \left(\hat{a}_{\bm{q},\lambda}^\dagger + \hat{a}_{-\bm{q}, \lambda}\right) + h.c. \Biggr]
    \end{aligned}
\end{equation}

This similarity, and indeed mutual mathematical interchangeability, of the exciton-exciton interaction provided by the Moiré potential and the structured optical modes is the central theoretical result of this work. In the following section we investigate the effect of $\hat{H}_{\textit{int}}$ on the excitonic system when the radiation field is treated both in a classical and quantum manner.

In our simulations we consider only the $1s$ MoSe$_2$ intra-layer exciton in the hetero-bilayer and reduce the photon space to the Fock number states $\{\ket{0}, \ket{1}\}$. This choice is motivated by computational feasibility but we expect the predicted phenomena to hold for inter-layer excitons at even larger grating periodicities.
Throughout this work, the cavity periodicity is always one dimensional (see Fig.~\ref{fig:cartoon}(d)).
Finally, we note that for the following comparison we either simulate the Moiré term $\mathcal{M}$ or the cavity periodicity $\mathcal{B}$, but we never consider both of them in the same simulation.

\begin{figure}[b!]
    \centering
    \begin{minipage}{0.49\linewidth}
        \vspace{-5mm}
        \fontsize{8}{8}\selectfont a) SF, $q_x = 0.000$ a.u., $\Omega_c = 1.468$ eV
        \centering
        \includegraphics[width=\linewidth]{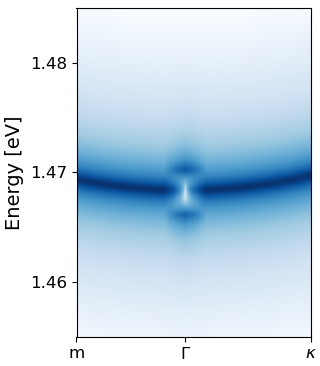}
    \end{minipage}
    \begin{minipage}{0.49\linewidth}
        \fontsize{8}{8}\selectfont b) LS, $q_x = 0.000$ a.u.
        \vspace{1mm}
        \centering
        \includegraphics[width=1.02\linewidth]{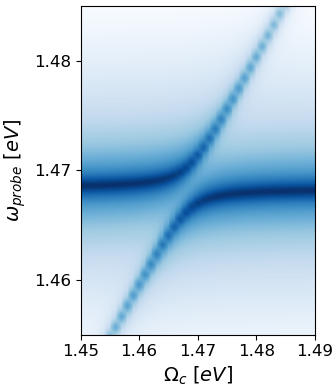}
    \end{minipage}
    \begin{minipage}{0.49\linewidth}
    \vspace{-0.2mm}
        \fontsize{8}{8}\selectfont c) SF, $q_x = \pm 0.009$ a.u., $\Omega_c = 1.463$ eV
        \centering
        \includegraphics[width=\linewidth]{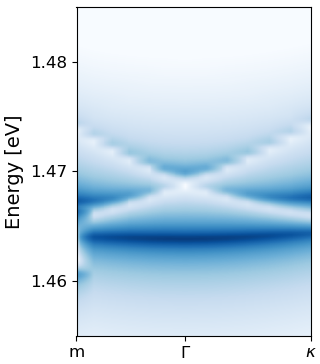}
    \end{minipage}
    \begin{minipage}{0.49\linewidth}
    \vspace{-0.2mm}
        \fontsize{8}{8}\selectfont d) SF, $q_x = \pm 0.009$ a.u., $\Omega_c = 0.05$ eV
        \centering
        \includegraphics[width=\linewidth]{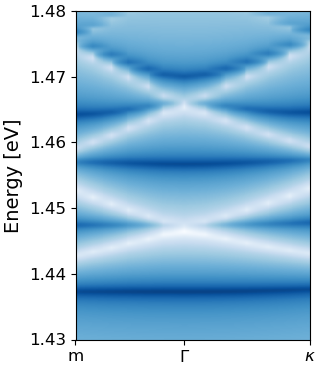}
    \end{minipage}
    \begin{minipage}{\linewidth}
        \vspace{2mm}
        \begin{flushright}
            \includegraphics[width=0.905\linewidth]{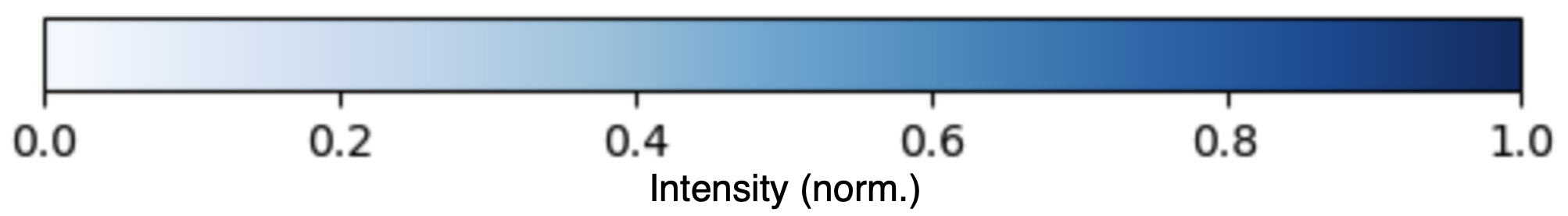}
        \end{flushright}
    \end{minipage}
    \caption{Spectral function (SF, c.f. Sec. \ref{sec:methods}, panels (a, c, d)) and imaginary part of the linear susceptibility (LS, c.f. Sec. \ref{sec:methods}, panel (b)) for the $1s$ intra-layer exciton in the MoSe$_2$ layer when light can be treated classically and in the absence of twist-induced Moiré potential ($\mathcal{M} = 0$ in Eq.~\ref{eq:h_interaction}). We use Floquet theory to solve the light-matter problem. We used $\tilde{A}_0 = 0.04$ a.u. for panels a-c and $\tilde{A}_0 = 0.02$ a.u. for panel (d). All panels use a normalized log scale for the intensity. For the spectral function, we use the standard $M-\Gamma-K$ path (which we shorten to $m-\Gamma-\kappa$, as shown in Fig.~\ref{fig:cartoon}(c) and Appendix~\ref{subsec:spectral_function}). (\textbf{a}, \textbf{b}) the light does not carry any momentum, so no terms in the Hamiltonian connect two \textit{k}-points. Hence, we observe a parabolic dispersion in the spectral function. In the linear response, we observe avoided crossing of the UP and LP formed from the excitonic state. (\textbf{c}, \textbf{d}) we used two photon modes carrying momentum $q_x = \pm 0.009$ a.u. In this case, the bilinear coupling generates a confining potential, and we observe the folding of the bands. Panel (c) is at excitonic resonance (we can observe the Rabi splitting). The value slightly differs from panel (a) due to the bi-linear coupling $\mathcal{B}$. On the contrary, panel (d) is at off-resonance, but the spectrum is rich due to the mixing of the Floquet replicas. The value of $\Omega_c = 0.05$ eV ensures the mixing of Floquet replicas.}
    \label{fig:classical_light}
\end{figure}

\begin{figure*}[t]
    \centering
    \begin{minipage}{0.49\textwidth}
        \fontsize{8}{8}\selectfont a) $\tilde{A}_0 = 0.0$ a.u.
        \centering
        \includegraphics[width=\linewidth]{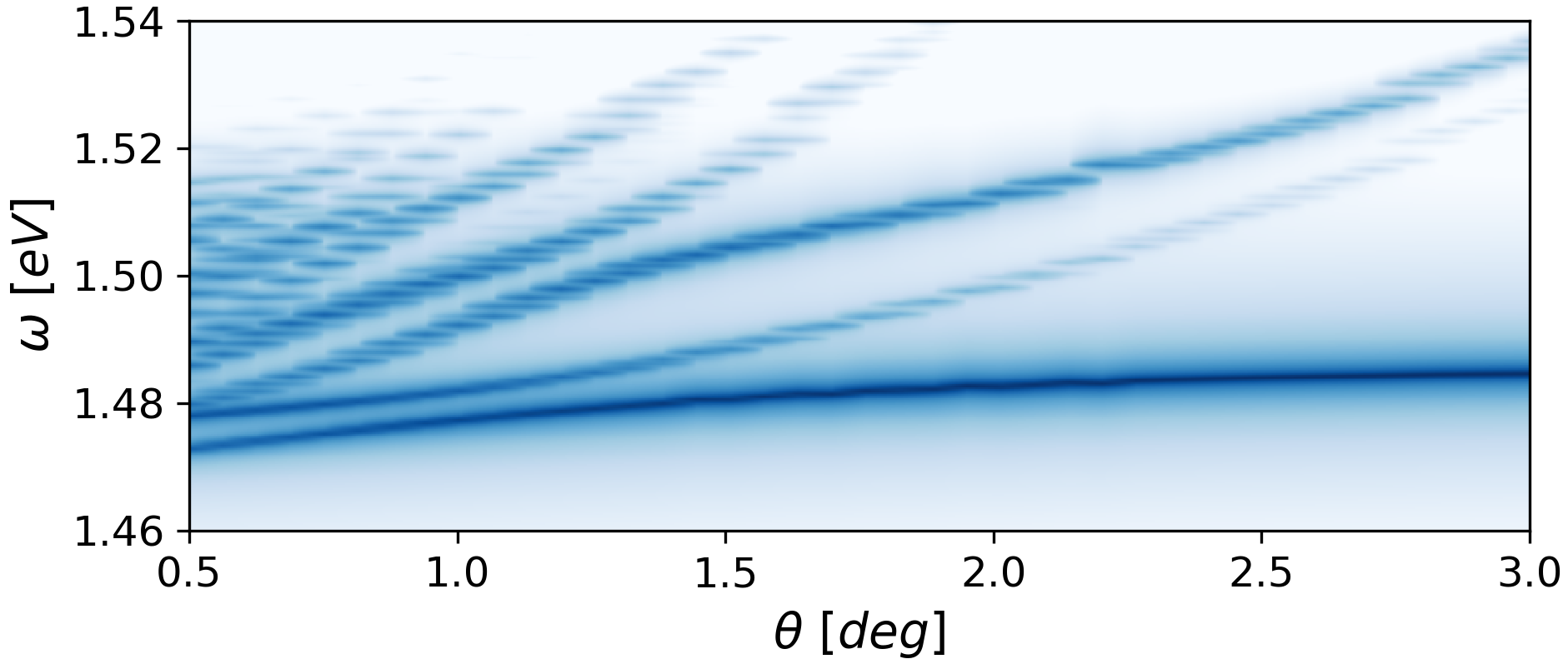}
    \end{minipage}
    \begin{minipage}{0.49\textwidth}
        \fontsize{8}{8}\selectfont b) $\tilde{A}_0 = 0.08$ a.u., $\Omega_c = 1.303$ eV
        \centering
        \includegraphics[width=\linewidth]{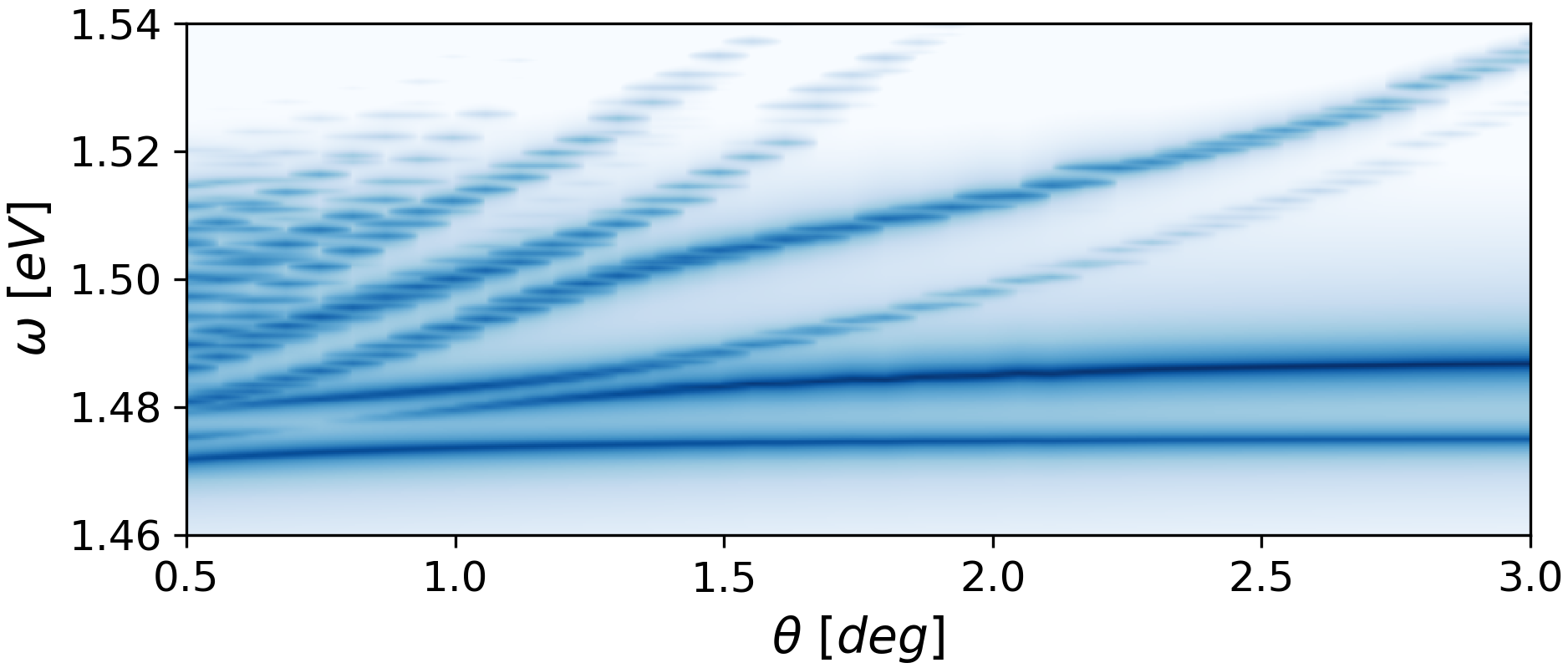}
    \end{minipage}
    \begin{minipage}{0.49\textwidth}
        \fontsize{8}{8}\selectfont c) $\tilde{A}_0 = 0.16$ a.u., $\Omega_c = 0.779$ eV
        \centering
        \includegraphics[width=\linewidth]{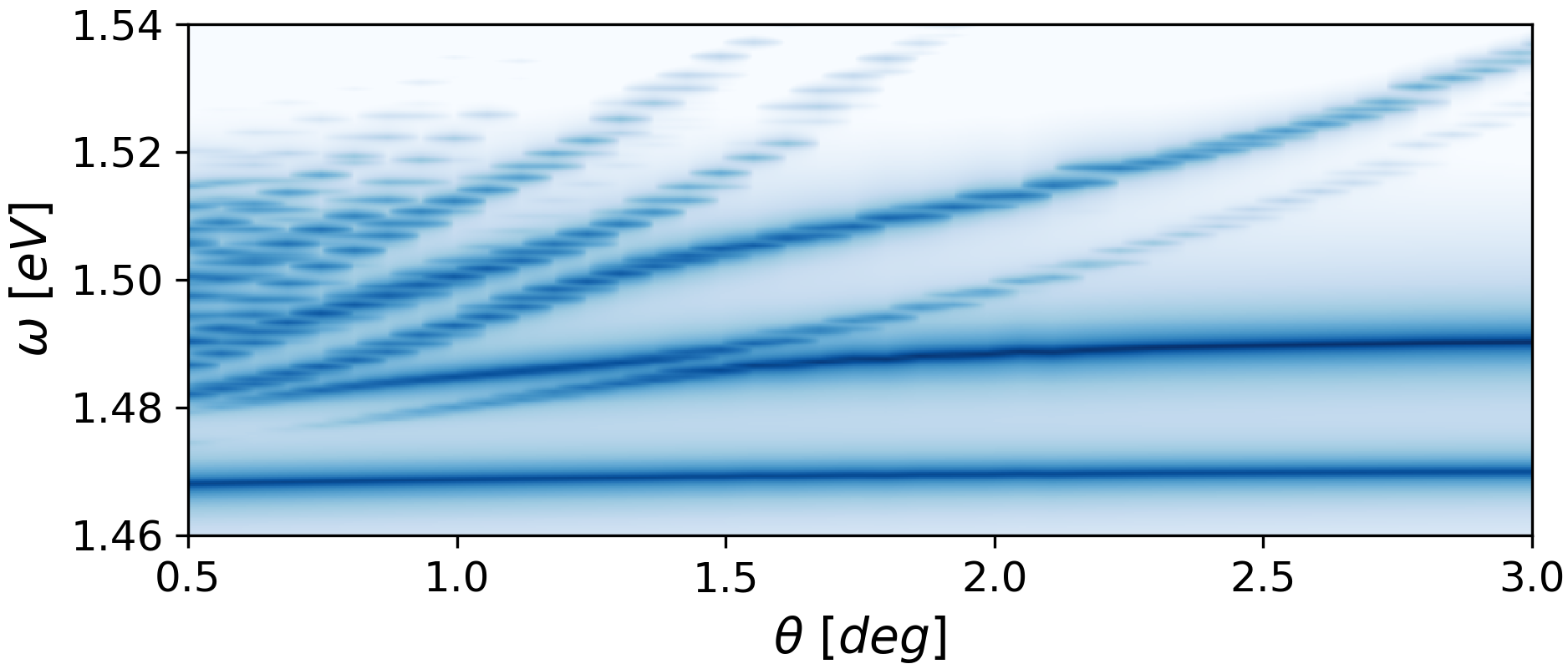}
    \end{minipage}
    \begin{minipage}{0.49\textwidth}
        \fontsize{8}{8}\selectfont d) UP and LP for the lowest branch
        \centering
        \includegraphics[width=\linewidth]{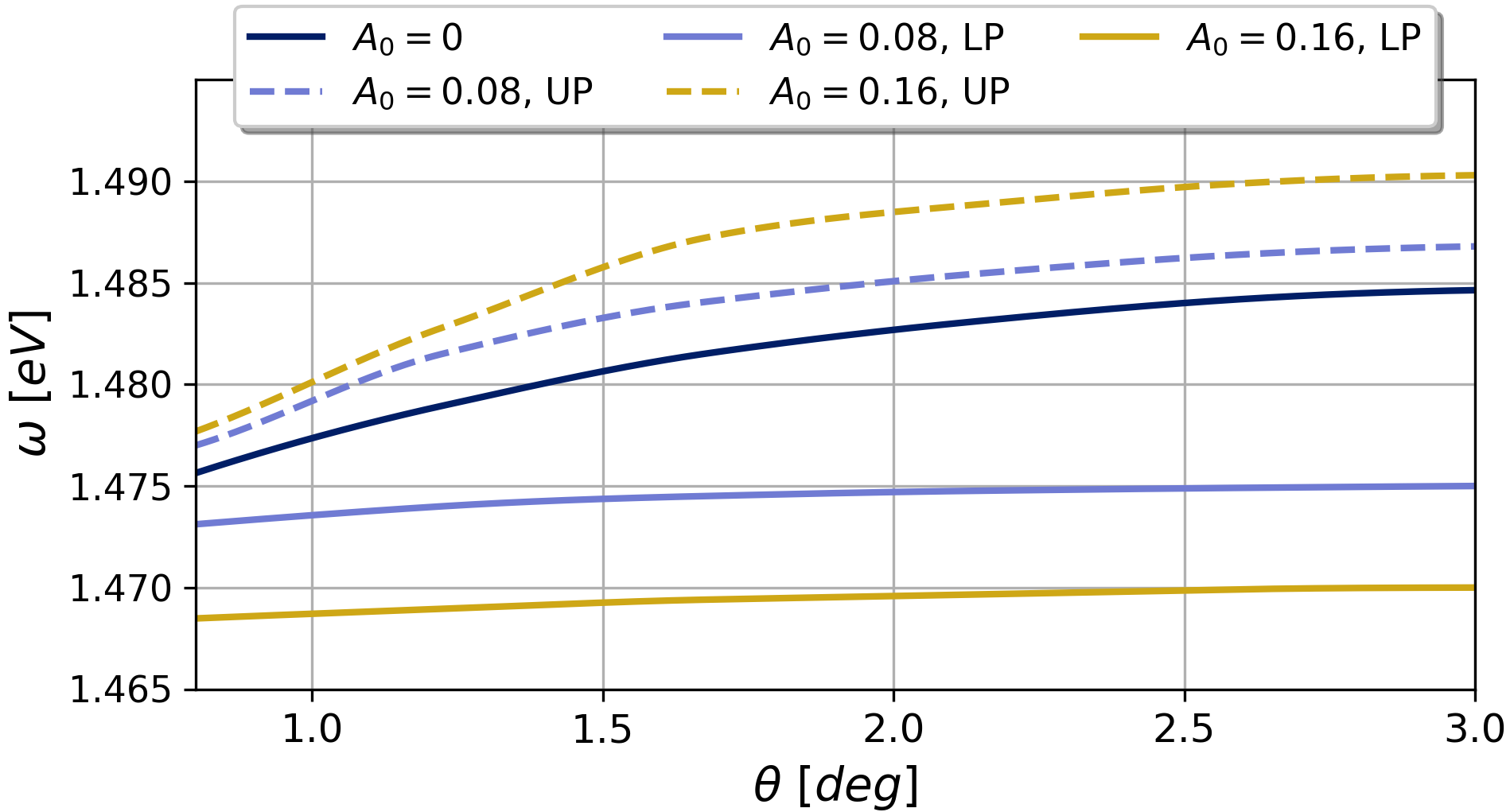}
    \end{minipage}
    \caption{Imaginary part of the linear susceptibility for the $1s$ intra-layer exciton in the MoSe$_2$ layer, as a function of the twist angle. The spacing used for the grid allows to resolve at most $\Delta \theta = 0.05$ deg, which explains the coarse nature of the lines (especially at small angles). Panels (a-c) use a normalized log scale for the intensity (see Fig.~\ref{fig:classical_light}). (\textbf{a}) Twisted hetero-structure without the cavity (see Ref.~\cite{tunable_phases} Fig. 5(a)). (\textbf{b}, \textbf{c}) Twisted hetero-structure embedded in a spatially unstructured cavity ($\bm{\Bar{q}} = \bm{0}$) for different light-matter coupling strengths. After fixing the energy of the mode to a constant value for all twist angles (in resonance with the lowest branch), we scanned over $\theta$. While all other states are mostly unaffected by the cavity, the lowest branch is split into the Upper (UP) and Lower (LP) Polariton, and the separation increases with the light-matter coupling $\tilde{A}_0$. Note that for these two panels $\Omega_c$ is significantly smaller than the excitonic resonance due to the diamagnetic shift. (\textbf{d}) Trace of the UP and LP formed from the lowest branch of the bare excitonic system (taken from the previous panels). The separation between UP and LP increases with the light-matter coupling. Furthermore, the LP is almost flat for all twist angles, meaning it is mainly unperturbed by the Moiré potential in the presence of a cavity, whereas the UP is affected by the Moiré potential only at low angles.}
    \label{fig:localization_lin_response}
\end{figure*}

\subsection{Classically driven cavity}
\label{subsec:classical_cavity}
We consider the effect of a driven structured cavity on the excitonic states of the untwisted bi-layer material (i.e. $\mathcal{M}=0$) and show that it gives confinement signatures similar to a twist induced Moiré potential. 
Examining Eq.~\ref{eq:h_interaction}, we note that the bi-linear coupling $\mathcal{B}$ shares the same excitonic operators of the Moiré potential $\mathcal{M}$, but it is additionally  paired with photonic operators.
In the case of classical radiation these operators are replaced by their mean-field value. To describe the driven system, we assume that the far field driving can couple to finite momentum cavity modes ($\pm\bm{\bar{q}}$) via the grating of the cavity. We model this excitation for single-polarization ($\lambda = s$) modes as a time-dependent coherent state $|\tilde{\lambda}_{\bm{\bar{q}}} \tilde{\lambda}_{\bm{-\bar{q}}} (t) \rangle = |\tilde{\lambda}_{\bm{\bar{q}}} (t)\rangle \otimes |\tilde{\lambda}_{\bm{-\bar{q}}} (t)\rangle$. Here we define the time-dependent coherent states as $|\tilde{\lambda}_{\bm{\bar{q}}} (t)\rangle=e^{-\frac{i \omega_{\bm{\bar{q}}}  t}{2}} |e^{-i \omega_{\bm{\bar{q}}} t} \tilde{\lambda}_{\bm{\bar{q}}}\rangle$ where the last ket goes by the usual definition $|\alpha\rangle = e^{-\frac{|\alpha|^2}{2}} \sum_{s=0}^{\infty} \frac{\tilde{\alpha}^s}{\sqrt{s!}}\ket{s}$ of coherent states. 
%\begin{equation}
%    |\tilde{\lambda}_{\bm{\bar{q}}} \tilde{\lambda}_{\bm{-\bar{q}}} t\rangle = e^{-\frac{i \left(\omega_{\bm{\bar{q}}} + \omega_{\bm{-\bar{q}}} \right) t}{2}} |e^{-i \omega_{\bm{\bar{q}}} t} \tilde{\lambda}_{\bm{\bar{q}}}, e^{-i \omega_{\bm{-\bar{q}}} t} \tilde{\lambda}_{\bm{-\bar{q}}}\rangle
%\end{equation}
%\[\hat{a}_{\bm{q}} \ket{\tilde{\lambda}_{\bm{q}}, \tilde{\lambda}_{\bm{-q}}, t} = \tilde{\lambda}_{\bm{q}} e^{-\frac{i \left(3\omega_{\bm{q}} + \omega_{\bm{-q}} \right) t}{2}} \ket{\tilde{\lambda}_{\bm{q}}, e^{-i \omega_{\bm{-q}} t} \tilde{\lambda}_{\bm{-q}}}\]
%where $|\tilde{\lambda}_{\bm{\bar{q}}} \tilde{\lambda}_{\bm{-\bar{q}}} t \rangle = |\tilde{\lambda}_{\bm{\bar{q}}}, t\rangle \otimes |\tilde{\lambda}_{\bm{-\bar{q}}}, t\rangle$ and $|\tilde{\lambda}_{\bm{\bar{q}}}\rangle = e^{-\frac{|\tilde{\lambda}_{\bm{\bar{q}}}|^2}{2}} \sum_{s=0}^{\infty} \frac{\tilde{\lambda}^s}{\sqrt{s!}}\ket{s_{\bm{\bar{q}}}}$.
Using that $\omega_{\bm{\bar{q}}} = \omega_{\bm{-\bar{q}}}$ and $\tilde{A}_{0, \bm{\bar{q}}} = \tilde{A}_{0, \bm{-\bar{q}}}$, and projecting the interaction Hamiltonian in Eq.~\ref{eq:h_interaction} onto these states introduces a time dependence of the form:
\[
    \langle \tilde{\lambda}_{\bm{\bar{q}}} \tilde{\lambda}_{\bm{-\bar{q}}} (t) | \hat{H}_{\text{int}} | \tilde{\lambda}_{\bm{\bar{q}}} \tilde{\lambda}_{\bm{-\bar{q}}} (t) \rangle \sim H^{(n=1)} e^{i \omega_{\bm{\bar{q}}} t} + H^{(n=-1)} e^{-i \omega_{\bm{\bar{q}}} t}
\]
where $H^{(n=\pm1)}$ is the time-\textit{independent} projection of the interaction Hamiltonian onto the coherent state, with $n$ the Floquet frequency index. See Appendix~\ref{app:h_classical} for the complete expression.

Since the obtained Hamiltonian is time periodic, we use Floquet theory to solve it~\cite{VanVleck, Hubener2021}.
The advantage of this approach is that the Floquet Hamiltonian is time independent, which makes it computationally cheaper while still being able to capture the polaritonic effects from the light-matter coupling.
With this semi-classical treatment light with finite momentum $\bm{\bar{q}}$ enters the coupled Hamiltonian in Eq.~\ref{eq:h_interaction} exactly as the potential created by a Moiré super-lattice, by generating interacting Floquet replicas, as sketched in Fig.~\ref{fig:cartoon}(b). This means that we can use classically driven cavities to induce an all-optical Moiré potential.

Fig.~\ref{fig:classical_light} shows the results for a classically driven unstructured cavity as well as for a spatially structured cavity (with a one-dimensional spatial periodicity) both coupled to the $1s$ intra-layer exciton in the MoSe$_2$ layer within the hetero-structure.
For the unstructured cavity (Fig.~\ref{fig:classical_light}(a-b)), where we study the system at the resonance between the cavity mode and the exciton, the dispersion relation of the exciton remains parabolic, as seen in the spectral function (SF, c.f. Section~\ref{sec:methods} for its definition) in Fig.~\ref{fig:classical_light}(a).
On top of the unperturbed dispersion, we observe a signature of the formation of an exciton-polariton by the appearance of a splitting at $\Gamma$, corroborated by the expected avoided crossing of upper and lower polariton branches in the imaginary part of the linear susceptibility (LS, c.f. Section~\ref{sec:methods} for its definition) as a function of cavity energies, our proxy for an absorption experiment (c.f. Fig.~\ref{fig:classical_light}(b)). 
%Note that since we are interested in showing the excitonic dispersion, the resolution in momentum  Fig.~\ref{fig:classical_light}a is too large to show the photon dispersion~\cite{Krol2019}, and this means that the photon line is not visible.
We then shift focus to a structured cavity both at and out of resonance with the excitons (Fig.~\ref{fig:classical_light}(c-d)). Here the excitons experience a confining potential generated by the bilinear coupling, which modifies the excitonic bands. Specifically, at resonance (Fig.~\ref{fig:classical_light}(c)) we observe a simple band folding stemming form the periodicity of the grating together with the mixing with the finite momentum replica of the many-body ground state. The polariton splitting is now shifted to finite momentum (the $m$-point), as compared to Fig.~\ref{fig:classical_light}(a). 
Off-resonance, at small frequencies (Fig.~\ref{fig:classical_light}(d)), the positive and negative excitonic replicas mix, generating a rich spectrum. We find that to achieve this mixing the cavity energy should be such that the excitonic replicas can intersect within the BZ.
Finally we find that at large off-resonant driving frequencies (not shown), the external driving does not induce any modification to the parabolic dispersion. This is because within the first-order high-frequency expansion of the Floquet Hamiltonian~\cite{VanVleck}, $H^{\omega \rightarrow\infty}_{\rm Floquet} = H^{(n=0)} + \frac{\left[H^{(n=-1)}, H^{(n=1)}\right]}{\Omega} + \mathcal{O}\left(\frac{1}{\Omega^2}\right)$, the commutator in the numerator vanishes, i.e. $\left[H^{(n=-1)}, H^{(n=1)}\right] = 0$.
In the Appendix~\ref{subsec:high_frequency} we show the explicit calculation of the commutator as well as the spectral function for various values of the cavity frequency.

\subsection{Dark cavity}
\label{subsec:dark_cavity}

\begin{figure*}[t]
    \centering
    \begin{minipage}{0.49\textwidth}
        \fontsize{8}{8}\selectfont a) SF for $\bm{\bar{q}_1} = [0.021, 0.0]$ a.u.; $\bm{\bar{q}_2} = [-0.021, 0.0]$ a.u.
        \centering
        \includegraphics[width=\textwidth]{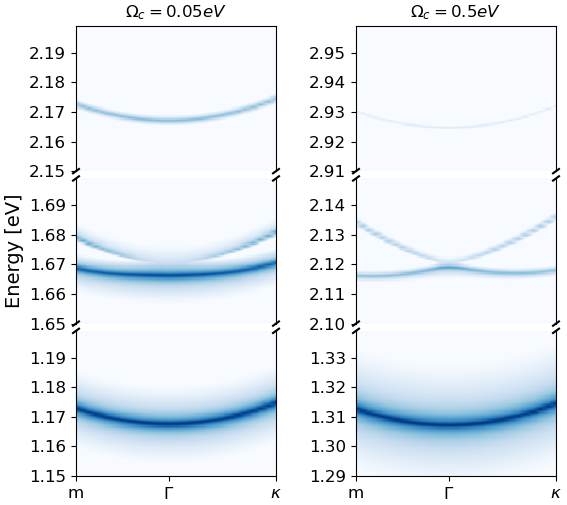}
        \vspace{0.5mm}
    \end{minipage}
    \begin{minipage}{0.49\textwidth}
        \fontsize{8}{8}\selectfont b) SF for $\bm{\bar{q}_1} = [0.033, 0.0]$ a.u.; $\bm{\bar{q}_2} = [-0.033, 0.0]$ a.u.
        \centering
        \includegraphics[width=\textwidth]{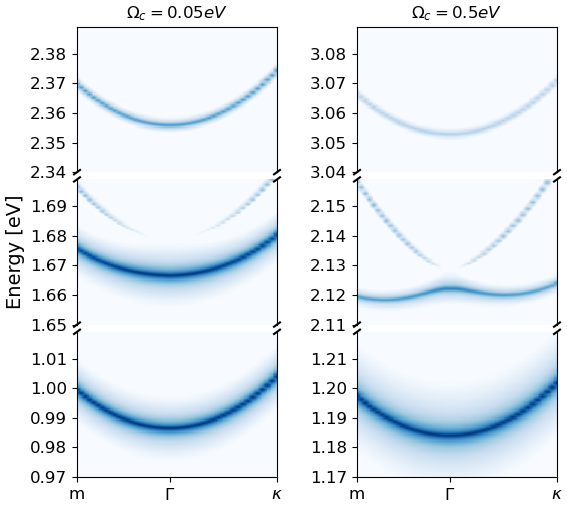}
        \vspace{0.5mm}
    \end{minipage}
    \caption{Spectral function for the untwisted hetero-structure ($\mathcal{M} = 0$ in Eq.~\ref{eq:h_interaction}) in a dark spatially structured optical cavity with two photonic modes. On the \textit{x}-axis we report the \textit{k}-point. \textit{m} and $\kappa$ are the borders of the BZ with periodicity induced by the momentum carried by light (see Appendix~\ref{subsec:spectral_function}). We used $\tilde{A}_{0, \bm{\bar{q}}} = 0.08$ a.u. for all panels. All panels use a normalized log scale for the intensity (see Fig.~\ref{fig:classical_light}). Note that the separation between the three bands depends both on the cavity energy $\Omega_c$, on the diamagnetic term $\tilde{A}_{0, \bm{\bar{q}}}^2$ and on the interaction energy $\tilde{A}_{0, \bm{\bar{q}}} \mathcal{B}_{ll',ij,\bm{Q \bar{q}}}^{\nu, \lambda}$. Finally, the position of the bottommost band is shifted towards lower energies due to the interaction energy contribution.}
    \label{fig:sf_finite_q}
\end{figure*}

We now analyze the cavity-exciton system for a dark cavity, i.e. in absence of external fields, where the effect of the photons on the matter is via their quantum fluctuations.
We begin by focusing on the quantum effects on the hetero-structure embedded in a spatially unstructured cavity ($\mathcal{B}=0$), including in this case the periodic confining potential from the Moiré pattern caused by twisting.
In the absence of momentum exchange, light and matter are coupled only through the bilinear coupling associated with $\mathcal{I}$ in Eq.~\ref{eq:h_interaction}.
Here, we examine how the unstructured planar cavity influences the twist-angle dependence of the excitonic response. For reference, in Fig.~\ref{fig:localization_lin_response}(a), we show the imaginary part of the linear susceptibility (c.f. Section~\ref{sec:methods} for its definition) as function of the twist angle $\theta$ and the optical probe energy $\omega$ outside the cavity.
As the twist angle decreases, the Moiré Brillouin zone contracts, the excitonic bands fold closer to the $\Gamma$ point, and more folded states enter the investigated energy window.
Therefore, the resulting linear response displays a richer spectrum at small twist angles, in line with earlier findings~\cite{tunable_phases}.

In the presence of cavity-matter coupling, when the cavity resonates with the MoSe$_2$ excitonic transition, we expect the Rabi splitting to dominate the behavior of the exciton-polariton response. We observe that compared to the case of uncoupled light and matter (Fig.~\ref{fig:localization_lin_response}(a)), the coupling splits the lowest branch into two, the upper (UP) and lower polariton (LP) (Fig.~\ref{fig:localization_lin_response}(b-c)). The magnitude of this splitting increases linearly with the light-matter coupling $\tilde{A}_0$.
Interestingly, the UP in Fig.~\ref{fig:localization_lin_response}(b-c) has the same dispersion as in the uncoupled matter system (Fig.~\ref{fig:localization_lin_response}(a)), whereas the LP line is almost constant for all twist angles.
This is shown in Fig.~\ref{fig:localization_lin_response}(d), where we traced out the curves obtained by the UP and the LP of Fig.~\ref{fig:localization_lin_response}(b-c) together with the lowest branch of the bare excitonic system (Fig.~\ref{fig:localization_lin_response}(a)).
A flat line in the spectrum means that the twist angle, which controls the size of the Moire Brillouin-Zone, has little influence on the excitonic states. This implies that the LP behaves as a low-energy unperturbed and unconfined exciton. 
Conversely, the dispersion of the UP is steeper and linear at small twist angles and then flattens out for larger angles, where it behaves similar to the LP, i.e. as nearly free particles. In the range of small twist angles, however, we can identify localized polaritonic states~\cite{tunable_phases}. 
In short, the effect of the twisting angle can be modified by the electromagnetic field of a cavity, which influences the excitonic response by causing Rabi splitting of the resonant transition. This results in a different localization for the upper and lower polariton.

\begin{figure}[t]
    \centering
    \begin{minipage}{0.49\linewidth}
        \fontsize{8}{8}\selectfont a) $\Omega_c = 0.05$ eV
        \centering
        \includegraphics[width=0.9\linewidth]{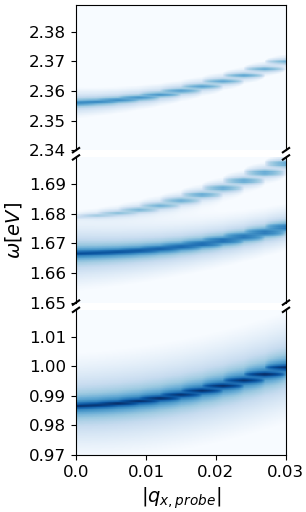}
    \end{minipage}
    \begin{minipage}{0.49\linewidth}
        \fontsize{8}{8}\selectfont b) $\Omega_c = 0.5$ eV
        \centering
        \includegraphics[width=0.9\linewidth]{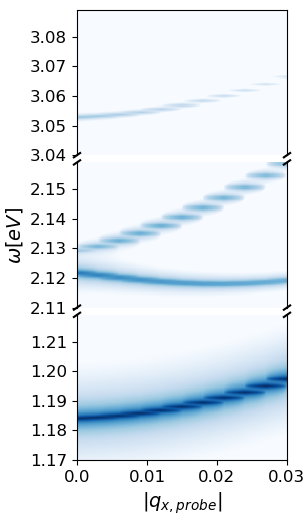}
    \end{minipage}
    \caption{Imaginary part of the linear susceptibility at finite $\bm{q}$ for the untwisted hetero-structure ($\mathcal{M} = 0$) in a dark spatially structured optical cavity with two photonic modes with $\bm{\bar{q}_1} = [0.033, 0.0]$ a.u.; $ \bm{\bar{q}_2} = [-0.033, 0.0]$ a.u.. We used $\tilde{A}_{0, \bm{q}} = 0.08$ a.u. for all panels. All panels use a normalized log scale for the intensity (see Fig.~\ref{fig:classical_light}). On the \textit{x}-axis we report the absolute value of the momentum carried by the probe beam, while on the \textit{y}-axis its energy. $|\bm{q}_{x, \rm{probe}}| = 0.01 \text{ a.u.}$ means that the probe is composed of two modes, one with $\bm{q_1} = [0.01, 0.0]$ a.u.; $\bm{q_2} = [-0.01, 0.0]$ a.u. This figure should be compared with Fig.~\ref{fig:sf_finite_q}(b). While the first and third bands maintain the parabolic dispersion regardless of the energy of the cavity, the central one visibly changes.}
    \label{fig:lin_res_finite_q}
\end{figure}

We then consider a spatially structured cavity, where both creation of excitons and Moiré-like exciton-scattering are included, i.e. both $\mathcal{B}$ and $\mathcal{I}$ are finite in Eq.~\ref{eq:h_qed_final}, while we let $\mathcal{M}$ = 0 to isolate the effect of the periodic grating of the cavity.
First, we analyze the results in Fig.~\ref{fig:sf_finite_q} which shows the excitonic spectral function. In all panels there are three different sets of bands.
From the analysis of the polaritonic states, resulting from the diagonalization of Eq.~\ref{eq:h_qed_final}, we identify the lowest energy sector to be mainly composed of an uncoupled excitonic state and the vacuum ($n=0$) state of both photonic modes. The central one is mainly constituted by an uncoupled excitonic state and one photon excitation ($n=1$) in either of the modes. Finally the high-energy band corresponds to an uncoupled excitonic state and a single excitation in both photon modes.
The separation between the three bands increases with the cavity energy $\Omega_c$, with the diamagnetic term $\tilde{A}_{0, \bm{\bar{q}}}^2$ and with the interaction energy $\tilde{A}_{0, \bm{\bar{q}}} \mathcal{B}_{ll',ij,\bm{Q \bar{q}}}^{\nu, \lambda}$.
Inspecting the uppermost and bottommost energy bands of Fig.~\ref{fig:sf_finite_q}, the curvature of the dispersion relation increases with the transfer of cavity photon momentum.
As the spectral function is related to the excitonic band structure, we can interpret this variation as the modification of the excitonic mass.
Since a steeper curvature corresponds to a smaller mass, the exciton becomes lighter as the photon momentum increases.
In the central band instead, the dispersion relation goes from parabolic to \textit{M}-shaped (i.e. the $\Gamma$ point is a local maximum instead of a global minimum). This implies that the excitonic mass is negative around this point, and might lead to the decay of the zero-momentum exciton towards one of the local minima, creating a new stable exciton-polariton state at finite momentum.
It should be noted that for the central band to show such features, the condition $\tilde{A}_{0, \bm{\bar{q}}} \mathcal{B}_{ll',ij,\bm{Q \bar{q}}}^{\nu, \lambda} \approx \tilde{A}_{0, \bm{-\bar{q}}} \mathcal{B}_{ll',ij,\bm{Q, -\bar{q}}}^{\nu, \lambda}$ should hold (i.e. both modes have the same interaction energy). Otherwise, the two exciton-polaritons would exist at different energy scales and could not scatter.
To prove that this effect could be experimentally demonstrated we show in Fig.~\ref{fig:lin_res_finite_q} the linear susceptibility as a function of probe photon momentum. In this figure going from left to right corresponds to going from $\Gamma$ to $m$ in Fig.~\ref{fig:sf_finite_q}(b).
As one can see, this observable is also able to capture the features discussed for the spectral function.

To better understand the physics, let us compare the effect of the bilinear coupling with that of the Moiré potential. In general, they both allow an exciton to scatter between different $k$-points. However, the different origin of the two terms plays a fundamental role in the effect on the excitonic system.
The Moiré potential $\mathcal{M}$ acts as a periodic scattering potential, which allows momentum transfer between excitons. 
Conversely, the \textit{optical} Moiré term $\mathcal{B}$, while also allowing the hopping of an exciton between $k$-points, it does so at the expense of creating/annihilating a cavity photon.
The underlying physics becomes even clearer when the excitons and the photons are off-resonant, so that photons can only be created/destroyed virtually. In this situation, the \textit{optical} Moiré term turns into a many-body, exciton-exciton interaction, whose momentum dependence is solely determined by the cavity design.
To explicitly show this feature, we make use of the high-frequency limit of the QED Hamiltonian. Within this limit, we can write a photon-free QED Hamiltonian by downfolding the original one in a dressed excitonic space according to~\cite{ferroelectric_gs, CavityGraphene, EffectiveLattice, BrillouinWigner}:
\begin{equation}
    \label{eq:h_downfolding}
    \begin{aligned}
        & \hat{H}^{\omega \rightarrow\infty}_{\rm QED} \approx \bra{0_{\bm{\bar{q}}}, 0_{\bm{-\bar{q}}}} \hat{H} \ket{0_{\bm{\bar{q}}}, 0_{\bm{-\bar{q}}}} - \\
        & \sum_{\bm{\bar{q}}} \frac{1}{\omega_{\bm{\bar{q}}}} \bigg[ \bra{0_{\bm{\bar{q}}}, 0_{\bm{-\bar{q}}}} \hat{H} \ket{1_{\bm{\bar{q}}}, 0_{\bm{-\bar{q}}}} \cdot \bra{1_{\bm{\bar{q}}}, 0_{\bm{-\bar{q}}}} \hat{H} \ket{0_{\bm{\bar{q}}}, 0_{\bm{-\bar{q}}}} \bigg]  .
    \end{aligned}
\end{equation}
Performing this expansion on the interaction Hamiltonian in Eq.~\ref{eq:h_interaction} leads to (see Appendix~\ref{app:downfolding} for the full derivation):
\begin{equation}
    \label{eq:h_high_frequency}
    \begin{aligned}
        & \hat{H}^{\omega \rightarrow\infty}_{\text{int}} = \sum_{ll', \nu} \sum_{i \in C, j \in V} \sum_{\bm{Q}} \mathcal{M}_{ll',ij,\bm{Q q}}^{\nu, \lambda} \hat{X}_{ll',ij,\bm{Q+q}}^{\nu \dagger} \hat{X}_{ll',ij,\bm{Q}}^{\nu} - \\
        & \frac{2 \tilde{A}_{0,\bm{q}} \tilde{A}_{0,\bm{-q}}}{\omega_{\bm{q}}} \sum_{ll' l_1 l_1'} \sum_{ij i_1 j_1} \sum_{\bm{Q} \bm{Q}_1, \nu \nu_1} \Biggl( \mathcal{B}_{ll',ij,\bm{Q, -q}}^{\nu, \lambda} \mathcal{B}_{l_1 l_1',i_1 j_1,\bm{Q_1 q}}^{\nu_1, \lambda} \\
        & \hat{X}_{ll',ij,\bm{Q-q}}^{\nu \dagger} \hat{X}_{l_1 l_1',i_1 j_1,\bm{Q_1+q}}^{\nu_1 \dagger} \hat{X}_{ll',ij,\bm{Q}}^{\nu} \hat{X}_{l_1 l_1',i_1 j_1,\bm{Q_1}}^{\nu_1} - \\
        & \left. \sum_{\bm{q}} \mathcal{I}_{ll',ij,\bm{Q q}}^{\nu, \lambda} \mathcal{I}_{l_1 l_1',i_1 j_1,\bm{Q_1 q}}^{\nu_1, \lambda, *} \hat{X}_{ll',ij,\bm{Q}}^{\nu \dagger} \hat{X}_{l_1 l_1',i_1 j_1,\bm{Q_1}}^{\nu_1} \right)
    \end{aligned}
\end{equation}
The striking result is the emergence of a four excitonic operator term, i.e. a term that describes an interaction between two excitons, which is a fundamental difference with respect to the Moiré potential.
This alters the excitonic dynamics as manifested by our calculated spectral functions and the non-trivial linear response features.
These excitons-excitons interactions, absent in classical treatments, highlight the potential of quantum cavities to engineer correlated excitonic phases.

%Despite the high-frequency expansion allowed us to identify this feature, we present the results for the full diagonalization of the multi-mode QED Hamiltonian (Fig \ref{fig:sf_finite_q} and \ref{fig:lin_res_finite_q}).

\subsection{Summary}\label{subsec:summary}
In this work, we demonstrated that spatially structured optical cavities can precisely emulate Moiré-like excitonic confinement in Van der Waals hetero-structures, circumventing the need for physical lattice twisting.

For classically light-driven cavities, momentum-carrying modes replicate the spatial modulation of a standard Moiré potential, leading to band folding and splitting, typical of a free particle in a periodic potential.
Such effects are predicted to be in the meV energy scale (for the parameters we used).

Conversely, in dark cavities the light-matter interaction is mediated solely by quantum vacuum fluctuations of the cavity modes, which induce long-range exciton-exciton interactions, fundamentally altering the excitonic dynamics. Note that we have recently observed that in Graphene cavity can induce long-range interaction for electrons~\cite{CavityGraphene}.
These interactions, not present when the system interacts with classical light, highlight the potential of quantum cavities to engineer correlated excitonic phases.
Most notably, we obtain a negative excitonic mass around the $\Gamma$ point that could be an experimental fingerprint of the cavity mediated excitonic interactions predicted in this work.

Our findings bridge the gap between twist engineering of two-dimensional hetero-structures and cavity quantum electrodynamics, providing a versatile platform for optically programmable excitonic systems.

%%%%%%%%%%%%%%%%%%%%%%%%%%%%%%%%%%%%%%%%%%%%%%%%%
% Methods
%%%%%%%%%%%%%%%%%%%%%%%%%%%%%%%%%%%%%%%%%%%%%%%%%
\section{Methods}\label{sec:methods}
In this work, we distinguish between spatially unstructured cavities, of the Fabry-Perot kind, where the electromagnetic field is treated with a single effective mode description in the long wavelength approximation~\cite{qed_hamiltonian} ($\bm{\bar{q}} = [q_x, q_y] = [0, 0]$), and spatially structured cavities, whose description requires two effective momentum carrying modes ($\bm{\bar{q}_1} = [q_x, 0], \bm{\bar{q}_2} = [-q_x, 0]$) corresponding to a cavity with a one-dimensional periodicity (like the one in Fig.~\ref{fig:cartoon}(d)). Based on symmetry arguments, we assume that $\omega_{\bm{\bar{q}_1}} = \omega_{\bm{\bar{q}_2}}$ and $\tilde{A}_{0, \bm{\bar{q}_1}} = \tilde{A}_{0, \bm{\bar{q}_2}}$.
Such photonic environments can be realized by means of dielectric metasurfaces~\cite{barton2020high, danielsen2025fourier}, or in polaritonic and plasmonic cavities~\cite{HerzigSheinfux2024, Galiffi2024, Chen2020}. In order to illustrate the broad potential of cavity-structuring of excitons, we leave the cavity mode energy, volume and grating wavelength as adjustable parameters.

In order to obtain the values of the modes momentum used in the previous sections ($\bm{\bar{q}_i} = [\pm 0.009, 0.0], [\pm 0.021, 0.0], [\pm 0.033, 0.0]$ a.u.), one should design the grating of the cavity accordingly. We estimate that the periodicity of the corresponding grating should be $d = \frac{2\pi}{\bm{\bar{q}_i}} = 40$ nm, $16$ nm, $10$ nm. We expect these extreme grating sizes to be larger for the optical confinement for inter-layer excitons.

In order to obtain the full exciton-polariton states, we represent the Hamiltonian on the $1s$ MoSe$_2$ intra-layer exciton and the Fock number states $\{\ket{0}, \ket{1}\}$ for the photonic modes.
The physics of the coupled light-matter system is then investigated by computing excitonic quantities: the linear optical susceptibility (LS) function and the spectral function (SF). 
The former represents the optical response of the system, obtained from applying linear response theory to the polaritonic states~\cite{Ruggenthaler2018}:
\begin{equation}
    \chi\left(\omega, \Omega_c, \theta\right) = \sum_I \frac{|\mathcal{M}_{I,0}|^2}{\omega - (E_I \left(\Omega_c, \theta\right) - E_0 \left(\Omega_c, \theta\right)) + i\eta}
\end{equation}
where $\mathcal{M}_{I,0}$ is the transition matrix element between the polaritonic ground state $\Psi_0$ and an excited state $\Psi_I$. $\eta$ is a small artificial broadening, $\Omega_c$ is the energy of the cavity modes, $\omega$ is the energy of the probe field, $\theta$ is the Moiré twist angle and $E_I \left(\Omega_c, \theta\right)$ is the energy of the $I$th polaritonic state.
Note that when we set $\mathcal{M} = 0$ (no standard Moiré potential), $\chi$ does not depend on $\theta$.
Furthermore, note that we only calculate the matter part of such a response by tracing out the photons.
To investigate the excitonic dispersion, i.e. their band structure, we employ the spectral function defined as:
\begin{equation}
    \mathcal{S}\left(\omega, \Omega_c, \theta, \bm{Q}\right) = \sum_I \frac{\bra{\Psi_I}\hat{X}_{\bm{Q}}^\dagger\ket{\Psi_0}\bra{\Psi_0}\hat{X}_{\bm{Q}}\ket{\Psi_I}}{\omega - (E_I \left(\Omega_c, \theta\right) - E_0 \left(\Omega_c, \theta\right)) + i\eta}
\end{equation}
This represents the probability of creating an exciton with an energy $E_I \left(\Omega_c, \theta\right)$ at a certain $k$-point.
Note that when we set $\mathcal{M} = 0$ (no standard Moiré potential), $\mathcal{S}$ does not depend on $\theta$.
Refer to the Appendix~\ref{app:observables} for additional insights.

%%%%%%%%%%%%%%%%%%%%%%%%%%%%%%%%%%%%%%%%%%%%%%%%%
% Acknowledgements
%%%%%%%%%%%%%%%%%%%%%%%%%%%%%%%%%%%%%%%%%%%%%%%%%
\begin{acknowledgments}
We acknowledge support from the Villum foundation grant No. 72146, the Cluster of Excellence "CUI: Advanced Imaging of Matter" - EXC 2056 - project ID 390715994, European Research Council (ERC-2024-SyG-101167294; UnMySt) and Grupos Consolidados (IT1453-22), and the Max Planck-New York City Center for Non-Equilibrium Quantum Phenomena. We acknowledge support from the European Union Marie Sklodowska-Curie Doctoral Networks TIMES grant No. 101118915 and SPARKLE grant No. 101169225. The Flatiron Institute is a division of the Simons Foundation.

\end{acknowledgments}

\bibliography{references}

\newpage

\appendix

\begin{widetext}
\section{QED Hamiltonian derivation} \label{app:qed_hamiltonian}
\noindent This section contains the derivation of the QED Hamiltonian.
We start by formulating it with electronic creation and annihilation operators, based on previous works \cite{tunable_phases, qed_hamiltonian}.
Subsequently, we change the basis, firstly to an electron-hole pair basis and finally to the exciton basis \cite{tunable_phases}.
Finally, we show that by doing so the Moiré potential and the bilinear coupling for finite $\bm{\Bar{q}}$ cavities enter the QED Hamiltonian in the same term.

\subsection{QED Hamiltonian with electronic creation and annihilation operators}
This work studies Moiré excitons in type-II MoSe$_2$/WSe$_2$ hetero-structure in optical cavities. Refer to Fig. 1 of the main text for its representation. The two layers are separated by a dielectric medium. Let us call $l$ the index of the layer.
Each TMD is represented with a set of $k$-points and a set of valence and conduction band states. An electron in one of the valence states of either layer is allowed to transition to any conduction state of the same or the other layer, resulting in an intra-layer or inter-layer exciton, respectively.

\noindent The uncoupled matter Hamiltonian of the system $\hat{H}_M$, which describes the hetero-structure, can be formulated as the sum of three different terms, the free particle, the Moiré and the Coulomb potential. Using the electronic creation and annihilation operators, the Hamiltonian reads:
\begin{equation}
    \label{eq:h_start}
    \begin{aligned}
        & \hat{H}_M = \sum_{l,i,\bm{k}}^C \varepsilon_{l,i,\bm{k}} \hat{c}_{l,i,\bm{k}}^\dagger \hat{c}_{l,i,\bm{k}} + \sum_{l,i,\bm{k}}^V \varepsilon_{l,i,\bm{k}} \hat{v}_{l,i,\bm{k}}^\dagger \hat{v}_{l,i,\bm{k}} + \sum_{ll'}\sum_{ij}^C\sum_{\bm{k} \bm{q}} V_l^C\left(\bm{q}\right) \hat{c}_{l,i,\bm{k+q}}^\dagger \hat{c}_{l',j,\bm{k}} + \\
        &   \sum_{ll'}\sum_{ij}^V\sum_{\bm{k} \bm{q}} V_l^V\left(\bm{q}\right) \hat{v}_{l,i,\bm{k+q}}^\dagger \hat{v}_{l',j,\bm{k}} + \sum_{ll'}\sum_{ij}\sum_{\bm{k} \bm{k}'\bm{q}} \mathcal{W}_{\bm{q}}^{ll'} \hat{c}_{l,i,\bm{k+q}}^\dagger \hat{v}_{l',j,\bm{k' - q}}^\dagger \hat{v}_{l',j,\bm{k}'} \hat{c}_{l,i,\bm{k}} + h.c.
    \end{aligned}
\end{equation}

where $i,j$ are band indexes which span over conduction ($C$) or valence ($V$) band states, $l,l'$ are layer indexes, $\bm{k}, \bm{k'}, \bm{q}$ are \textit{k}-points indexes. The operators $\hat{c}^\dagger, \hat{c}$ ($\hat{v}^\dagger, \hat{v}$) create or annihilate an electron in the conduction (valence) band.
$\mathcal{W}_{\bm{q}}^{ll'}$ is the matrix element of the Coulomb potential \cite{2d_coulomb}.
$V_l \left(\bm{q}\right)$ is the Moiré potential coefficient \cite{tunable_phases}:

\begin{equation}
    \label{eq:moire_prefactor}
    V_l\left(\bm{q}\right) = v_l\sum_{n=0}^2 e^{iC_3^n\left(\bm{G_l^0}+\bm{G_{l'}^0})\right) \cdot \bm{D_l} / 2} \delta_{\bm{q},C_3^n\left(\bm{G_l^0} - \bm{G_{l'}^0})\right)}; v_l = \alpha_l + e^{2\pi i \sigma_{l'}/3}\beta_l
\end{equation}
The values of $\alpha_l, \beta_l$ are taken from Table 1 of the Supporting Information of Ref. \cite{tunable_phases}, while $\sigma_l = 1$ for all layers (R-type stacking).

We describe the uncoupled light system with an Hamiltonian consisting of a set of effective harmonic oscillators. As for the light-matter coupling, we perform the canonical momentum substitution $\bm{\hat{p}} \rightarrow \bm{\hat{p} + \hat{A}}$ on the uncoupled matter Hamiltonian, obtaining the second-quantized Pauli-Fierz Hamiltonian \cite{qed_hamiltonian}.
The full QED Hamiltonian can then be formulated as follows \cite{qed_hamiltonian}:

\begin{equation}
    \label{eq:h_qed_single_particle_operators}
    \begin{aligned}
        \hat{H}_{QED} = \hat{H}_M + \hat{H}_b + \hat{H}_d + \sum_{\bm{\bar{q}},\lambda} \omega_{\bm{\bar{q}}} \left(\hat{a}_{\bm{\bar{q}}, \lambda}^\dagger \hat{a}_{\bm{\bar{q}}, \lambda} + \frac{1}{2}\right)
    \end{aligned}
\end{equation}

where $\lambda$ is the polarization of the mode and $\hat{a}$ ($\hat{a}^{\dagger}$) is the annihilation (creation) operator for the mode $\ket{\bm{\bar{q}}, \lambda}$. $\omega_{\bm{\bar{q}}}$ is the energy of the mode $\bm{\Bar{q}}$. Moreover, we define the renormalized light-matter coupling constant as $\tilde{A}_{0,\bm{\bar{q}}} = \frac{A_{0, \bm{\bar{q}}}}{\sqrt{V_{\textit{eff}, \bm{\bar{q}}}}}$ ($A_{0, \bm{\bar{q}}}$ is the coupling strength of the mode $\bm{\bar{q}}$ and $V_{\textit{eff}, \bm{\bar{q}}}$ is the effective mode volume).
$\hat{H}_b$ and $\hat{H}_d$, the bilinear and diamagnetic Hamiltonians, are:

\begin{equation}
    \label{eq:h_bil_single_particle_operators}
    \begin{aligned}
        \hat{H}_{b} = \sum_{\lambda} \sum_{\bm{k},\bm{\Bar{q}}} \Tilde{A}_{0,\bm{\Bar{q}}} \left[ \sum_{ij}^C p_{ij,\bm{k+\Bar{q}},\bm{k}}^\lambda \hat{c}_{i,\bm{k+\Bar{q}}}^\dagger \hat{c}_{j,\bm{k}} \right. + \sum_{i,j}^V p_{ij,\bm{k+\Bar{q}},\bm{k}}^\lambda \hat{v}_{i,\bm{k+\Bar{q}}}^\dagger \hat{v}_{j,\bm{k}} + \left. \sum_{i \in C, j \in V} p_{ij,\bm{k+\Bar{q}},\bm{k}}^\lambda \hat{c}_{i,\bm{k+\Bar{q}}}^\dagger \hat{v}_{j,\bm{k}} \right] \left(\hat{a}_{\bm{\Bar{q}},\lambda}^\dagger + \hat{a}_{-\bm{\Bar{q}}, \lambda}\right)
    \end{aligned}
\end{equation}

\begin{equation}
    \label{eq:h_dia_single_particle_operators}
    \begin{aligned}
        & \hat{H}_{d} = \sum_{\lambda \lambda'} \sum_{\bm{k} \bm{\Bar{q}} \bm{\Bar{q}}'} \frac{\Tilde{A}_{0,\bm{\Bar{q}}} \Tilde{A}_{0,\bm{\Bar{q}}'}}{2} \left(\hat{a}_{\bm{\Bar{q}},\lambda}^\dagger + \hat{a}_{-\bm{\Bar{q}}, \lambda}\right) \left(\hat{a}_{\bm{\Bar{q}}',\lambda'}^\dagger + \hat{a}_{-\bm{\Bar{q}}', \lambda'}\right) * \\
        & \Biggl[ \sum_{ij}^C s_{ij,\bm{k-\Bar{q}+\Bar{q}'}}^{\lambda \lambda'} \hat{c}_{i,\bm{k+\Bar{q}}}^\dagger \hat{c}_{j,\bm{k}} +
        \sum_{ij}^V s_{ij,\bm{k-\Bar{q}+\Bar{q}'}}^{\lambda \lambda'} \hat{v}_{i,\bm{k-\Bar{q}+\Bar{q}'}}^\dagger \hat{v}_{j,\bm{k}} + 
        \sum_{i \in C, j \in V} s_{ij,\bm{k-\Bar{q}+\Bar{q}'}}^{\lambda \lambda'} \hat{c}_{i,\bm{k-\Bar{q}+\Bar{q}'}}^\dagger \hat{v}_{j,\bm{k}} \Biggl]
    \end{aligned}
\end{equation}

where $p_{ij,\bm{k+\Bar{q}},\bm{k}}^\lambda$ is the momentum matrix element and $s_{ij,\bm{k-\Bar{q}+\Bar{q}'}}^{\lambda \lambda'}$ is the overlap matrix element \cite{qed_hamiltonian}.

Note that in equations \ref{eq:h_qed_single_particle_operators}-\ref{eq:h_dia_single_particle_operators} the symbol $\bm{\Bar{q}}$ appears as the momentum associated with the photonic mode.
This shall not be confused with the $\bm{q}$ which appears in eq \ref{eq:h_start}-\ref{eq:moire_prefactor}, which is the reciprocal lattice vector associated with the periodicity of the Moiré potential.

\subsection{QED Hamiltonian in the electron-hole basis}
\noindent We prefer to use an exciton representation to study the behavior of excitons in such a system, as it allows to directly encode the effect of the Coulomb potential in the formation of such bound quasi-particles.
To address this problem, let us define the electron-hole operators and their low-density expansions \cite{tunable_phases}:
\begin{align}
    \label{eq:el_ho_op_create}
    & \hat{P}_{l,i,\bm{k};l',j,\bm{k}'}^\dagger = \hat{c}_{l,i,\bm{k}}^\dagger \hat{v}_{l',j,\bm{k}'} \\
    \label{eq:el_ho_op_annihilate}
    & \hat{P}_{l,i,\bm{k};l',j,\bm{k}'} = \hat{v}_{l',j,\bm{k}'}^\dagger \hat{c}_{l,i,\bm{k}} \\
    & \hat{c}_{l,i,\bm{k}}^\dagger \hat{c}_{l',j,\bm{k}'} \approx \sum_{m,a,\bm{q}} \hat{P}_{l,i,\bm{k};a,m,\bm{q}}^\dagger \hat{P}_{l',j,\bm{k}';m,a,\bm{q}} \\
    & \hat{v}_{l,i,\bm{k}}^\dagger \hat{v}_{l',j,\bm{k}'} \approx \delta_{ll',ij,\bm{k} \bm{k}'} - \sum_{m,a,\bm{q}} \hat{P}_{a,m,\bm{q};l,i,\bm{k}}^\dagger \hat{P}_{m,a,\bm{q};l',j,\bm{k}'}
\end{align}
In the following, the indexes of the $\hat{P}$ operators will always follow this order: layer index, band index, $\bm{k}$ point index. This is regardless of the letters used.
By using these relations, we can transform the QED Hamiltonian as follows:
\begin{equation}
    \label{eq:h_matter_el_hole_free}
    \begin{aligned}
        \hat{H}_{f} = \sum_{l,i,\bm{k}}^V \varepsilon_{l,i,\bm{k}} - \sum_{l,j,\bm{k}}^V \varepsilon_{l,j,\bm{k}} \sum_{r,s,\bm{p}}^C \hat{P}_{r,s,\bm{p};l,j,\bm{k}}^\dagger \hat{P}_{r,s,\bm{p},l,j,\bm{k}} + \sum_{l,i,\bm{k}}^C \varepsilon_{l,i,\bm{k}} \sum_{r,s,\bm{p}}^V \hat{P}_{l,i,\bm{k};r,s,\bm{p}}^\dagger \hat{P}_{l,i,\bm{k};r,s,\bm{p}}
    \end{aligned}
\end{equation}

\begin{equation}
    \label{eq:h_matter_el_hole_moire}
    \begin{aligned}
        \hat{H}_{m} = \sum_{l,j,\bm{k}}^V V_{l,i}\left(\bm{0}\right) - \sum_{ij}^V \sum_{ll'} \sum_{\bm{kq}} V_{ll',ij}\left(\bm{q}\right) \sum_{r,s,\bm{p}}^C \hat{P}_{r,s,\bm{p};l',j,\bm{k}}^\dagger \hat{P}_{r,s,\bm{p};l,i,\bm{k+q}} + \sum_{ij}^C \sum_{ll'} \sum_{\bm{kq}} V_{ll',ij}\left(\bm{q}\right) \sum_{r,s,\bm{p}}^V \hat{P}_{l,i,\bm{k+q};r,s,\bm{p}}^\dagger \hat{P}_{l,i,\bm{k};r,s,\bm{p}}
    \end{aligned}
\end{equation}

\begin{equation}
    \label{eq:h_matter_el_hole_coulomb}
    \begin{aligned}
        \hat{H}_{c} = \sum_{l,i,\bm{k}}^C \mathcal{W}_{\bm{0}}^{ll'} \hat{c}_{l,i,\bm{k}}^\dagger \hat{c}_{l,i,\bm{k}} - \sum_{l,i,\bm{k}}^C \sum_{l',j,\bm{k}'}^V \sum_{\bm{q}} \mathcal{W}_{\bm{q}}^{ll'} \hat{P}_{l,i,\bm{k+q};l',j,\bm{k}'}^\dagger \hat{P}_{l,i,\bm{k+q};l',j,\bm{k'-q}}
    \end{aligned}
\end{equation}

\begin{equation}
    \label{eq:h_bil_el_hole_operators}
    \begin{aligned}
        & \hat{H}_{bil} = \sum_{ll'} \sum_{\bm{k},\bm{\bm{\Bar{q}}},\lambda} \Tilde{A}_{0,\bm{\Bar{q}}} \Biggl[ \sum_{ij}^V p_{ij,ll',\bm{k+\Bar{q},k}}^\lambda \delta_{ij,ll',\bm{\Bar{q}=0}} + \sum_{ij}^C \sum_{r,s,\bm{p}}^V p_{ij,\bm{k+\Bar{q}},\bm{k}}^\lambda \hat{P}_{l,i,\bm{k+\Bar{q}};r,s,\bm{p}}^\dagger \hat{P}_{l',j,\bm{k};r,s,\bm{p}} - \\
        & \sum_{ij}^V \sum_{r,s,\bm{p}}^C p_{ij,\bm{k+\Bar{q}},\bm{k}}^\lambda \hat{P}_{r,s,\bm{p};l',j,\bm{\Bar{q}}}^\dagger \hat{P}_{r,s,\bm{p};l,i,\bm{k+\Bar{q}}} + \sum_{i \in C, j \in V} p_{ij,\bm{k}+\bm{\Bar{q}}, \bm{k}}^\lambda \hat{P}_{l,i,\bm{k + \Bar{q}};l',j,\bm{k}}^\dagger + h.c. \Biggl] \left(\hat{a}_{\bm{\Bar{q}},\lambda}^\dagger + \hat{a}_{-\bm{\Bar{q}}, \lambda}\right)
    \end{aligned}
\end{equation}

\begin{equation}
    \label{eq:h_dia_el_hole_operators}
    \begin{aligned}
        & \hat{H}_{dia} = \sum_{\lambda \lambda'} \sum_{ll'} \sum_{\bm{k},\bm{\Bar{q}},\bm{\Bar{q}}'} \frac{\Tilde{A}_{0,\bm{\Bar{q}}} \Tilde{A}_{0,\bm{\Bar{q}}'}}{2} \Biggl[ \sum_{ij}^V s_{ij,ll',\bm{k-\Bar{q}+\Bar{q}'}}^{\lambda \lambda'} \delta_{ij,ll',\bm{\Bar{q}=\Bar{q}'=0}} + \sum_{ij}^C \sum_{r,s,\bm{p}}^V s_{ij,\bm{k-\Bar{q}+\Bar{q}'}}^{\lambda \lambda'} \hat{P}_{l,i,\bm{k-\Bar{q}+\Bar{q}'};r,s,\bm{p}}^\dagger \hat{P}_{l',j,\bm{k};r,s,\bm{p}} - \\
        &   \sum_{ij}^V \sum_{r,s,\bm{p}}^C s_{ij,\bm{k-\Bar{q}+\Bar{q}'}}^{\lambda \lambda'} \hat{P}_{r,s,\bm{p};l',j,\bm{k}}^\dagger \hat{P}_{r,s,\bm{p};l,i,\bm{k-\Bar{q}+\Bar{q}'}} + \sum_{i \in C, j \in V} s_{ij,\bm{k-\Bar{q}+\Bar{q}'}}^{\lambda \lambda'} \hat{P}_{l,i,\bm{k-\Bar{q}+\Bar{q}'};l',j,\bm{k}}^\dagger \Biggl] \left(\hat{a}_{\bm{\Bar{q}},\lambda}^\dagger + \hat{a}_{-\bm{\Bar{q}}, \lambda}\right) \left(\hat{a}_{\bm{\Bar{q}}',\lambda'}^\dagger + \hat{a}_{-\bm{\Bar{q}}', \lambda'}\right)
    \end{aligned}
\end{equation}

\subsection{QED Hamiltonian in the excitonic basis} \label{subsec:qed_c}
\noindent The basis used in the previous section can be further optimized for the present problem.
For this purpose, we introduce the bound excitons operators $\hat{X}_{\bm{Q}}^{ll'}$, where $\bm{Q}$ is the center of mass momentum of the exciton, such that:
\begin{equation}
    \label{eq:x_eq}
    \hat{P}_{l,i,\bm{k};l',j,\bm{k}'}^\dagger = \sum_{\nu} \hat{X}_{ll',\bm{k - k'}}^{\nu \dagger} \psi_{ll'}^{\nu} \left(\alpha_{ll'}\bm{k'} +\beta_{ll'}\bm{k}\right)
\end{equation}
where $\alpha_{ll'}$ and $\beta_{ll'}$ are the reduced electron and hole masses \cite{tunable_phases}, $\nu$ is the excitonic state (i.e. $\nu=1s, 2s...$) and $\psi$ is the excitonic wave function from the Wannier equation \cite{wannier}:
\begin{equation}
    \label{eq:wannier}
    \frac{\hbar^2 k^2}{2 m_{ll'}^r} \psi_{ll'}^{\nu}\left(\bm{k}\right) - \sum_{\bm{q}} \mathcal{W}_{\bm{q}}^{ll'} \psi_{ll'}^{\nu}\left(\bm{k + q}\right) = \hat{E}_{ll'}^{\nu} \psi_{ll'}^{\nu}\left(\bm{k}\right)
\end{equation}
In the following derivations, we will consider the $\nu$ index explicitly. However, in the simulation we limit to $\nu=1s$. In the previous equation, $\mathcal{W}_{\bm{q}}^{ll'}$ is the Coulomb potential defined in \ref{eq:h_start} and $\hat{E}_{ll'}^{\nu}$ is the binding energy.
Note that the Wannier equation gives the excitonic states for a specific combination of $l,l'$. In this work, solving it for $l=l'=1$ gives the intra-layer excitons for the $MoSe_2$ layer, while solving for $l=l'=2$ gives the intra-layer excitons for the $WSe_2$ layer and $l \neq l'$ gives the inter-layer excitons (electron jumping from a valence band state of $MoSe_2$ to the conduction band of $WSe_2$).

First let us see how the Hamiltonian (eq \ref{eq:h_matter_el_hole_free}-\ref{eq:h_dia_el_hole_operators}) transforms with the introduction of the $\hat{X}$ operators (eq \ref{eq:x_eq}). Moreover, we also define $\bm{Q} = \bm{k-p}$, which is convenient to simplify the equations.
Finally, we neglect the constant terms (i.e. the ones without any operator) in Eq. \ref{eq:h_matter_el_hole_free} and Eq. \ref{eq:h_matter_el_hole_moire}, as they only appear in the main diagonal of the matrix. As for the constant term in Eq. \ref{eq:h_bil_el_hole_operators}, we approximate $p_{ij,ll',\bm{k+\Bar{q},k}}^\lambda \delta_{ij,ll',\bm{\Bar{q}=0}} = p_{ii,ll,\bm{k,k}}^\lambda \approx \frac{\norm{\bm{k}}}{m_h}$ \cite{cavity_control}.
\begin{equation}
    \label{eq:h_matter_free_x_op}
    \begin{aligned}
        & \hat{H}_{f} = \sum_{\nu} \sum_{l,i,\bm{Q}}^C \sum_{r,s,\bm{k}}^V \varepsilon_{l,i,\bm{k}} \psi_{lr,is}^{\nu} \left[\alpha_{lr,is}\left(\bm{k-Q}\right) + \beta_{lr,is}\bm{k}\right] \psi_{lr,is}^{\nu *} \left[\alpha_{lr,is}\left(\bm{k-Q}\right) + \beta_{lr,is}\bm{k}\right] \hat{X}_{lr,is,\bm{Q}}^{\nu \dagger} \hat{X}_{lr,is,\bm{Q}} - \\
        &   \sum_{\nu} \sum_{l,j,\bm{Q}}^V \sum_{r,s,\bm{k}}^C \varepsilon_{l,j,\bm{k}} \psi_{ls,ir}^{\nu} \left[\alpha_{lr,is}\bm{k} + \beta_{lr,is}\left(\bm{k-Q}\right)\right] \psi_{ls,ir}^{\nu *} \left[\alpha_{lr,is}\bm{k} + \beta_{lr,is}\left(\bm{k-Q}\right)\right] \hat{X}_{lr,js,\bm{-Q}}^{\nu \dagger} \hat{X}_{lr,is,\bm{-Q}}
    \end{aligned}
\end{equation}

\begin{equation}
    \label{eq:h_matter_moire_x_op}
    %\begin{aligned}
    %    & \hat{H}_{m} = \sum_{ll',ij,\bm{Qq}}^C V_{ll',ij}\left(\bm{q}\right) \sum_{r,s,\bm{k}}^V \hat{X}_{lr,is,\bm{Q+q}}^{\nu \dagger} \psi_{lr,is}^{\nu} \left[\alpha_{lr}^{is}\left(\bm{k-Q}\right) + \beta_{lr}^{is}\left(\bm{k+q}\right)\right] \psi_{l'r,js}^{\nu *} \left[\alpha_{l'r}^{js}\left(\bm{k-Q}\right) + \beta_{l'r}^{js}\left(\bm{k}\right)\right] \hat{X}_{l'r,js,\bm{Q}} - \\
    %    &   \sum_{ll',ij,\bm{Qq}}^V V_{ll',ij}\left(\bm{q}\right) \sum_{r,s,\bm{k}}^C \hat{X}_{l'r,js,\bm{-Q}}^{\nu \dagger} \psi_{l'r,js}^{\nu} \left[\alpha_{l'r}^{js}\left(\bm{k}\right) + \beta_{l'r}^{js}\left(\bm{k-Q}\right)\right] \psi_{lr,is}^{\nu *} \left[\alpha_{lr}^{is}\left(\bm{k+q}\right) + \beta_{lr}^{is}\left(\bm{k-Q}\right)\right] \hat{X}_{lr,is,\bm{-Q+q}}
    %\end{aligned}
    \begin{aligned}
        & \hat{H}_{m} = \sum_{ll'} \sum_{i \in C, j \in V} \sum_{\bm{Q,q}, \nu} \left[\sum_{s}^C V_{ll'}^{is}\left(\bm{q}\right) \sum_{r, \bm{k}} \psi_{lr,ij}^{\nu} \left[\alpha_{lr}^{ij}\left(\bm{k-Q}\right) + \beta_{lr}^{ij}\left(\bm{k+q}\right)\right] \psi_{l'r,js}^{\nu *} \left[\alpha_{l'r}^{js}\left(\bm{k-Q}\right) + \beta_{l'r}^{js}\left(\bm{k}\right)\right] - \right. \\
        & \left. \sum_{s}^V V_{ll'}^{js}\left(\bm{q}\right)  \sum_{r, \bm{k}} \psi_{l'r,is}^{\nu} \left[\alpha_{l'r}^{is}\left(\bm{k}\right) + \beta_{l'r}^{is}\left(\bm{k-Q}\right)\right] \psi_{lr,is}^{\nu *} \left[\alpha_{lr}^{ij}\left(\bm{k+q}\right) + \beta_{lr}^{ij}\left(\bm{k-Q}\right)\right] \right] \hat{X}_{ll',ij,\bm{Q+q}}^{\nu \dagger} \hat{X}_{ll',ij,\bm{Q}}^{\nu}
    \end{aligned}
\end{equation}

\begin{equation}
    \label{eq:h_matter_coulomb_x_op}
    \begin{aligned}
        \hat{H}_{c} = \sum_{l,i,\bm{Q}}^C \sum_{l',j,\bm{k}}^V \sum_{\nu,\bm{q}} \mathcal{W}_{\bm{q}}^{ll'} \psi_{ll',ij}^{\nu} \left[\alpha_{ll'}^{ij}\left(\bm{k-Q}\right) + \beta_{ll'}^{ij}\left(\bm{k+q}\right)\right] \psi_{ll',ij}^{\nu *} \left[\alpha_{ll'}^{ij}\left(\bm{k-Q}\right) + \beta_{ll'}^{ij}\left(\bm{k+q}\right)\right] \hat{X}_{ll',ij,\bm{Q+q}}^{\nu \dagger} \hat{X}_{ll',ij,\bm{Q+q}}
    \end{aligned}
\end{equation}

\begin{equation}
    \label{eq:h_bil_x_op}
    %\begin{aligned}
    %    & \hat{H}_{bil} = \sum_{\lambda} \sum_{\bm{Q,\Bar{q}}} \sum_{ll'} \Tilde{A}_{0,\bm{\Bar{q}}} \Biggl[ \sum_{i}^V \frac{\norm{\bm{Q}}}{m_h} \delta_{ll',\bm{\Bar{q}=0}} + \\
    %    &   \sum_{ij}^C \sum_{r,s,\bm{k}}^V \sum_{\nu} p_{ij,\bm{k+\Bar{q}},\bm{k}}^\lambda \hat{X}_{lr,is,\bm{Q+\Bar{q}}}^{\nu \dagger} \psi_{lr,is}^{\nu} \left[\alpha_{lr,is}\left(\bm{k-Q}\right) + \beta_{lr,is}\left(\bm{k+\Bar{q}}\right)\right] \psi_{l'r,js}^{\nu *} \left[\alpha_{l'r,js}\left(\bm{k-Q}\right) + \beta_{l'r,js}\bm{k}\right] \hat{X}_{l'r,js,\bm{Q}}^\nu - \\
    %    & \sum_{ij}^V \sum_{r,s,\bm{k}}^C \sum_{\nu} p_{ij,\bm{k+\Bar{q}},\bm{k}}^\lambda \hat{X}_{l'r,js,\bm{-Q}}^{\nu \dagger} \psi_{l'r,js}^{\nu} \left[\alpha_{l'r,js}\bm{k} + \beta_{l'r,js}\left(\bm{k-Q}\right)\right] \psi_{lr,is}^{\nu *} \left[\alpha_{lr,is}\left(\bm{k+q}\right) + \beta_{lr,is}\left(\bm{k-Q}\right)\right] \hat{X}_{lr,is,-\bm{Q+\Bar{q}}}^\nu \\
    %    & \left. \sum_{i \in C, j \in V} \sum_\nu p_{ij,\bm{k}+\bm{\Bar{q}}, \bm{k}}^\lambda \hat{X}_{ll',ij,\bm{\Bar{q}}}^{\nu \dagger} \psi_{ll',ij}^{\nu} \left[\alpha_{ll'}^{ij}\left(\bm{k}\right) + \beta_{ll'}^{ij}\left(\bm{k-Q}\right)\right] + h.c. \right] \left(\hat{a}_{\bm{\Bar{q}},\lambda}^\dagger + \hat{a}_{-\bm{\Bar{q}}, \lambda}\right)
    %\end{aligned}
    \begin{aligned}
        & \hat{H}_{bil} = \sum_{\lambda, \bm{\bar{q}}} \tilde{A}_{0,\bm{\bar{q}}} \sum_{ll'} \sum_{i \in C, j \in V} \sum_{\bm{Q}, \nu} \left\{ \frac{\norm{\bm{Q}}}{m_h} \delta_{ll',ij,\bm{\bar{q}=0}} + \sum_{\bm{k}} p_{ij,\bm{k}+\bm{\bar{q}}, \bm{k}}^\lambda \psi_{ll',ij}^{\nu} \left[\alpha_{ll'}^{ij}\left(\bm{k}\right) + \beta_{ll'}^{ij}\left(\bm{k-Q}\right)\right] \hat{X}_{ll',ij,\bm{\bar{q}}}^{\nu \dagger} \right. + \\
        & \hat{X}_{ll',ij,\bm{Q+\bar{q}}}^{\nu \dagger} \hat{X}_{ll',ij,\bm{Q}}^{\nu} \left[ \sum_{r,s,\bm{k}}^C p_{is,\bm{k+\bar{q}},\bm{k}}^\lambda  \psi_{lr,ij}^{\nu} \left[\alpha_{lr}^{ij}\left(\bm{k-Q}\right) + \beta_{lr}^{ij}\left(\bm{k+\bar{q}}\right)\right] \psi_{l'r,js}^{\nu *} \left[\alpha_{l'r}^{js}\left(\bm{k-Q}\right) + \beta_{l'r}^{js}\left(\bm{k}\right)\right] - \right. \\
        & \left. \sum_{r,s,\bm{k}}^V p_{js,\bm{k+\bar{q}},\bm{k}}^\lambda \psi_{l'r,js}^{\nu} \left[\alpha_{l'r}^{js}\left(\bm{k}\right) + \beta_{l'r}^{js}\left(\bm{k-Q}\right)\right] \psi_{lr,ij}^{\nu *} \left[\alpha_{lr}^{ij}\left(\bm{k+\bar{q}}\right) + \beta_{lr}^{ij}\left(\bm{k-Q}\right)\right] \right] + h.c. \Biggl\} \left(\hat{a}_{\bm{\bar{q}},\lambda}^\dagger + \hat{a}_{-\bm{\bar{q}}, \lambda}\right)
    \end{aligned}
\end{equation}

\begin{equation}
    \label{eq:h_dia_x_op}
    %\begin{aligned}
    %    & \hat{H}_{dia} = \sum_{\lambda \lambda'} \sum_{l l'} \sum_{\bm{Q},\bm{\Bar{q}},\bm{\Bar{q}}'} \frac{\Tilde{A}_{0,\bm{\Bar{q}}} \Tilde{A}_{0,\bm{\Bar{q}}'}}{2} \Biggl[ \sum_{i}^V s_{ii,ll',\bm{k-\Bar{q}+\Bar{q}'}}^{\lambda, \lambda'} \delta_{ll',\bm{\Bar{q}=\Bar{q}'=0}} + \\
    %    &   \sum_{ij}^C \sum_{r,s,\bm{k}}^V s_{ij,\bm{k-\Bar{q}+\Bar{q}'}}^{\lambda \lambda'} \hat{X}_{lr,is,\bm{Q-\Bar{q}+\Bar{q}'}}^{\nu \dagger} \psi_{lr,is}^{\nu} \left[\alpha_{lr}^{is}\left(\bm{k-Q}\right) + \beta_{lr}^{is}\left(\bm{k-\Bar{q}+\Bar{q}'}\right)\right] \psi_{l'r,js}^{\nu *} \left[\alpha_{l'r}^{js}\left(\bm{k-Q}\right) + \beta_{l'r}^{js}\left(\bm{k}\right)\right] \hat{X}_{l'r,js,\bm{Q}}^\nu \\
    %    &   \sum_{ij}^V \sum_{r,s,\bm{k}}^C s_{ij,\bm{k-\Bar{q}+\Bar{q}'}}^{\lambda \lambda'} \hat{X}_{l'r,js,\bm{-Q}}^{\nu \dagger} \psi_{l'r,js}^{\nu} \left[\alpha_{l'r}^{js}\left(\bm{k}\right) + \beta_{l'r}^{js}\left(\bm{k-Q}\right)\right] \psi_{lr,is}^{\nu *} \left[\alpha_{lr}^{is}\left(\bm{k-\Bar{q}+\Bar{q}'}\right) + \beta_{lr}^{is}\left(\bm{k-Q}\right)\right] \hat{X}_{lr,is,\bm{-Q+\Bar{q}-\Bar{q}'}}^\nu \\
    %    &   \left. \sum_{i \in C, j \in V} s_{ij,\bm{k-\Bar{q}+\Bar{q}'}}^{\lambda \lambda'} \hat{X}_{ll',ij,\bm{\Bar{q}-\Bar{q}'}}^{\nu \dagger} \psi_{ll',ij}^{\nu} \left[\alpha_{ll'}^{ij}\left(\bm{k}\right) + \beta_{ll'}^{ij}\left(\bm{k-\Bar{q}+\Bar{q}'}\right)\right] + h.c. \right] \left(\hat{a}_{\bm{\Bar{q}},\lambda}^\dagger + \hat{a}_{-\bm{\Bar{q}}, \lambda}\right) \left(\hat{a}_{\bm{\Bar{q}}',\lambda'}^\dagger + \hat{a}_{-\bm{\Bar{q}}', \lambda'}\right)
    %\end{aligned}
    \begin{aligned}
        & \hat{H}_{dia} = \sum_{\lambda \lambda'} \sum_{\bm{\bar{q}},\bm{\bar{q}}'} \frac{\tilde{A}_{0,\bm{\bar{q}}} \tilde{A}_{0,\bm{\bar{q}}'}}{2} \left\{ \sum_{ll'} \sum_{i \in C, j \in V} \sum_{\bm{Q}, \nu} s_{ll,jj,\bm{Q}}^{\lambda \lambda'} \delta_{ij,ll',\bm{\bar{q}=\bar{q}'=0}} + \right.\\
        &   \left[ \sum_{r,s,\bm{k}}^C s_{is,\bm{k-\bar{q}+\bar{q}'}}^{\lambda \lambda'} \psi_{lr,ij}^{\nu} \left[\alpha_{lr}^{ij}\left(\bm{k-Q}\right) + \beta_{lr}^{ij}\left(\bm{k-\bar{q}+\bar{q}'}\right)\right] \psi_{l'r,js}^{\nu *} \left[\alpha_{l'r}^{js}\left(\bm{k-Q}\right) + \beta_{l'r}^{js}\left(\bm{k}\right)\right] -\right. \\
        &   \left. \sum_{r,s,\bm{k}}^V s_{js,\bm{k-\bar{q}+\bar{q}'}}^{\lambda \lambda'} \psi_{l'r,is}^{\nu} \left[\alpha_{l'r}^{is}\left(\bm{k}\right) + \beta_{l'r}^{ij}\left(\bm{k-Q}\right)\right] \psi_{lr,ij}^{\nu *} \left[\alpha_{lr}^{ij}\left(\bm{k-\bar{q}+\bar{q}'}\right) + \beta_{lr}^{is}\left(\bm{k-Q}\right)\right] \right] \hat{X}_{ll',ij,\bm{Q-\bar{q}+\bar{q}'}}^{\nu \dagger} \hat{X}_{ll',ij,\bm{Q}}^{\nu \dagger} + \\
        &   \left. \sum_{ll'} \sum_{i \in C, j \in V} \sum_{\bm{k}, \nu} s_{ij,\bm{k-\bar{q}+\bar{q}'}}^{\lambda \lambda'} \psi_{ll',ij}^{\nu} \left[\alpha_{ll'}^{ij}\left(\bm{k}\right) + \beta_{ll'}^{ij}\left(\bm{k-\bar{q}+\bar{q}'}\right)\right] \hat{X}_{ll',ij,\bm{\bar{q}-\bar{q}'}}^{\nu \dagger} + h.c. \right\} \left(\hat{a}_{\bm{\bar{q}},\lambda}^\dagger + \hat{a}_{-\bm{\bar{q}}, \lambda}\right) \left(\hat{a}_{\bm{\bar{q}}',\lambda'}^\dagger + \hat{a}_{-\bm{\bar{q}}', \lambda'}\right)
    \end{aligned}
\end{equation}
After having expressed the Hamiltonian in terms of the operators $\hat{X}$ and of the Wannier wavefunctions $\psi$, we can further simplify the expression by combining $\hat{H}_f$ and $\hat{H}_c$ using Eq. \ref{eq:wannier}.
Then, we can write:
\begin{equation}
    \label{eq:free_exciton}
    \hat{H}_{f} = \sum_{ll'}\sum_{i \in C, j \in V} \sum_{\nu,\bm{Q}} \mathcal{E}_{ll',ij,\bm{Q}}^{\nu} \hat{X}_{ll',ij,\bm{Q}}^{\nu \dagger} \hat{X}_{ll',ij,\bm{Q}}^{\nu}
\end{equation}

As a final step, for the sake of shortening the above expressions, it is convenient to define some common quantities. First, let us define:
\begin{equation}
    \label{eq:form_factor_psi}
    \Psi_{ll'r,ijs}^{\nu} \left(\bm{k}',\bm{k}'',\bm{k}''', \bm{k}''''\right) = \psi_{lr,ij}^{\nu} \left[\alpha_{lr}^{ij}\left(\bm{k}'\right) + \beta_{lr}^{ij}\left(\bm{k}''\right)\right] \psi_{l'r,is}^{\nu *} \left[\alpha_{lr}^{is}\left(\bm{k}'''\right) + \beta_{lr}^{is}\left(\bm{k}''''\right)\right]
\end{equation}
And subsequently the form factor:
\begin{equation}
    \label{eq:form_factor}
    \mathcal{F}_{ll',ijs}^{\nu} \left(\bm{k}',\bm{k}'',\bm{k}''', \bm{k}''''\right) = \sum_{r,\bm{k}} \Psi_{ll'r,ijs}^{\nu} \left(\bm{k}',\bm{k}'',\bm{k}''', \bm{k}''''\right)
\end{equation}
Note that the expression for the form factor is an extension of Eq. 16 of the Supplementary Information of \cite{tunable_phases}. The two equations become the same if we use only one valence and one conduction band to describe the system (i.e. if we drop the indexes $i, j, s$).
Using Eq. \ref{eq:form_factor} we can now define the Moiré potential prefactor as:
\begin{equation}
    \label{eq:moire_prefactor_form_factor}
    \mathcal{M}_{ll',ij,\bm{Q,q}}^{\nu} = \sum_{s}^{C} V_{ll'}^{is}\left(\bm{q}\right) \mathcal{F}_{ll',ijs}^{\nu} \left(\bm{k-Q}, \bm{k+q}, \bm{k-Q}, \bm{q}\right) - \sum_{s}^{V} V_{ll'}^{js}\left(\bm{q}\right) \mathcal{F}_{ll',ijs}^{\nu} \left(\bm{k}, \bm{k+q}, \bm{k}, \bm{k-Q}\right)
\end{equation}
After defining this quantity, we can rewrite Eq. \ref{eq:h_matter_moire_x_op} as:
\begin{equation}
    \label{eq:h_matter_moire_final}
    \begin{aligned}
        & \hat{H}_{m} = \sum_{ll'} \sum_{i \in C, j \in V} \sum_{\bm{Q,q}, \nu} \mathcal{M}_{ll',ij,\bm{Q,q}}^{\nu} \hat{X}_{ll',ij,\bm{Q+q}}^{\nu \dagger} \hat{X}_{ll',ij,\bm{Q}}^{\nu}
    \end{aligned}
\end{equation}
We can follow a similar strategy for both the bilinear coupling $\hat{H}_{bil}$ and the diamagnetic $\hat{H}_{dia}$ term. For the former we can define:
\begin{equation}
    \label{eq:b_coeff}
    \mathcal{B}_{ll',ij,\bm{Q,\Bar{q}}}^{\nu, \lambda} = \sum_{r,\bm{k}} \biggl[ \sum_{s}^{C} p_{is, \bm{k+\Bar{q},\bm{k}}}^{\lambda} \Psi_{ll'r, ijs}^{\nu} \left(\bm{k-Q}, \bm{k+\Bar{q}, \bm{k-Q}, \bm{k}}\right) - \sum_{s}^{V} p_{js, \bm{k+\Bar{q},\bm{k}}}^{\lambda} \Psi_{ll'r, ijs}^{\nu *} \left(\bm{k}, \bm{k-Q}, \bm{k+\Bar{q}}, \bm{k-Q}\right) \biggl]
\end{equation}
\begin{equation}
    \label{eq:i_coeff}
    \mathcal{I}_{ll',ij,\bm{Q,\Bar{q}}}^{\nu, \lambda} = \sum_{\bm{k}} p_{ij,\bm{k}+\bm{\Bar{q}}, \bm{k}}^\lambda \psi_{ll',ij}^{\nu} \left[\alpha_{ll'}^{ij}\left(\bm{k}\right) + \beta_{ll'}^{ij}\left(\bm{k-Q}\right)\right]
\end{equation}
While for the diamagnetic term one has that:
\begin{equation}
    \begin{aligned}
        \label{eq:d_coeff}
        & \mathcal{D}_{ll',ij,\bm{Q,\Bar{q},\Bar{q}'}}^{\nu, \lambda \lambda'} = \sum_{r,\bm{k}} \biggl[ \\
        & \sum_{s}^{C} s_{is, \bm{k-\Bar{q}+\Bar{q}'}}^{\lambda \lambda'} \Psi_{ll'r, ijs}^{\nu} \left(\bm{k-Q}, \bm{k-\Bar{q}+\Bar{q}', \bm{k-Q}, \bm{k}}\right) - \sum_{s}^{V} s_{js, \bm{k-\Bar{q}+\Bar{q}'}}^{\lambda \lambda'} \Psi_{ll'r, ijs}^{\nu *} \left(\bm{k}, \bm{k-Q}, \bm{k-\Bar{q}+\Bar{q}'}, \bm{k-Q}\right) \biggl]
    \end{aligned}
\end{equation}
\begin{equation}
    \label{eq:s_coeff}
    \mathcal{S}_{ll',ij,\bm{\Bar{q},\Bar{q}'}}^{\nu, \lambda \lambda'} = \sum_{\bm{k}} s_{ij, \bm{k-\Bar{q}+\Bar{q}'}}^{\nu, \lambda \lambda'} \psi_{ll',ij}^{\nu} \left[\alpha_{ll'}^{ij}\left(\bm{k}\right) + \beta_{ll'}^{ij}\left(\bm{k-\Bar{q}+\Bar{q}'}\right)\right]
\end{equation}
Finally, we can substitute Eq. \ref{eq:b_coeff}, \ref{eq:i_coeff}, \ref{eq:d_coeff}, \ref{eq:s_coeff} into Eq. \ref{eq:h_matter_moire_x_op}, \ref{eq:h_bil_x_op}, \ref{eq:h_dia_x_op} to obtain the full Hamiltonian in the excitonic base.
Before writing the final expression, we neglect the terms that appear with a delta in Eq. \ref{eq:h_bil_x_op}, \ref{eq:h_dia_x_op}, as the are just a constant.
%We neglect them as they do not allow for new transitions (as they are not bound to any excitonic operator) and their magnitude is rather small compared to the other elements in the bilinear coupling or diamagnetic term. In fact, they are not summed over $\bm{k}$.
Thus, the final Hamiltonian reads:
\begin{equation}
    \label{eq:h_full_x_op}
    \begin{aligned}
        & \hat{H}_{QED} = \sum_{\bm{\Bar{q}},\lambda} \omega_{\bm{\Bar{q}}} \left(\hat{a}_{\bm{\Bar{q}}, \lambda}^\dagger \hat{a}_{\bm{\Bar{q}}, \lambda} + \frac{1}{2}\right) + \sum_{ll'}\sum_{i \in C, j \in V} \sum_{\nu,\bm{Q}} \left( \mathcal{E}_{ll',ij,\bm{Q}}^{\nu} \hat{X}_{ll',ij,\bm{Q}}^{\nu \dagger} \hat{X}_{ll',ij,\bm{Q}}^{\nu} + \sum_{\bm{q}} \mathcal{M}_{ll',ij,\bm{Q,q}}^{\nu} \hat{X}_{ll',ij,\bm{Q+q}}^{\nu \dagger} \hat{X}_{ll',ij,\bm{Q}}^{\nu} \right) + \\
        & \sum_{\lambda, \bm{\Bar{q}}} \Tilde{A}_{0,\bm{\Bar{q}}} \sum_{ll'} \sum_{i \in C, j \in V} \sum_{\bm{Q}, \nu} \left[ \mathcal{I}_{ll',ij,\bm{Q,\Bar{q}}}^{\nu, \lambda} \hat{X}_{ll',ij,\bm{\Bar{q}}}^{\nu \dagger}  + \mathcal{B}_{ll',ij,\bm{Q,\Bar{q}}}^{\nu, \lambda} \hat{X}_{ll',ij,\bm{Q+\Bar{q}}}^{\nu \dagger} \hat{X}_{ll',ij,\bm{Q}}^{\nu} + h.c. \right] \left(\hat{a}_{\bm{\Bar{q}},\lambda}^\dagger + \hat{a}_{-\bm{\Bar{q}}, \lambda}\right) + \\
        &  \sum_{\lambda \lambda'} \sum_{\bm{\Bar{q}},\bm{\Bar{q}}'} \frac{\Tilde{A}_{0,\bm{\Bar{q}}} \Tilde{A}_{0,\bm{\Bar{q}}'}}{2} \sum_{ll'} \sum_{i \in C, j \in V} \sum_{\bm{Q}, \nu} \mathcal{D}_{ll',ij,\bm{Q,\Bar{q},\Bar{q}'}}^{\nu, \lambda, \lambda'} \hat{X}_{ll',ij,\bm{Q-\Bar{q}+\Bar{q}'}}^{\nu \dagger} \hat{X}_{ll',ij,\bm{Q}}^{\nu} \left(\hat{a}_{\bm{\Bar{q}},\lambda}^\dagger + \hat{a}_{-\bm{\Bar{q}}, \lambda}\right) \left(\hat{a}_{\bm{\Bar{q}}',\lambda'}^\dagger + \hat{a}_{-\bm{\Bar{q}}', \lambda'}\right) + \\
        & \sum_{\lambda \lambda'} \sum_{\bm{\Bar{q}},\bm{\Bar{q}}'} \frac{\Tilde{A}_{0,\bm{\Bar{q}}} \Tilde{A}_{0,\bm{\Bar{q}}'}}{2} \sum_{ll'} \sum_{i \in C, j \in V} \sum_{\bm{k}, \nu} \mathcal{S}_{ll',ij,\bm{\Bar{q},\Bar{q}'}}^{\nu, \lambda \lambda'} \hat{X}_{ll',ij,\bm{\Bar{q}-\Bar{q}'}}^{\nu \dagger} \left(\hat{a}_{\bm{\Bar{q}},\lambda}^\dagger + \hat{a}_{-\bm{\Bar{q}}, \lambda}\right) \left(\hat{a}_{\bm{\Bar{q}}',\lambda'}^\dagger + \hat{a}_{-\bm{\Bar{q}}', \lambda'}\right) + h.c.
    \end{aligned}
\end{equation}
Note that in the main text we absorb the diamagnetic term into the uncoupled photon Hamiltonian by performing a Bogoliubov transformation~\cite{CavityGraphene}.

\section{Observables} \label{app:observables}
This section discusses the formulas for computing the observables shown in the main text. As stated in the Results Section of the main text, we compute both the linear response function $\chi\left(\omega, \Omega_c, \theta\right)$ and the spectral function $\mathcal{S}\left(\omega, \Omega_c, \theta, \bm{Q}\right)$.

\subsection{Linear response function}
\label{subsec:linear_response}
The linear response function $\chi\left(\omega, \Omega_c, \theta\right)$ represents the optical response of the system. It is obtained from applying the linear response theory to the polaritonic states \cite{Ruggenthaler2018}. We only calculate the matter part of such a response by tracing out the photons. 
Hence, we formulate it as:
\begin{equation}
    \label{eq:linear_response}
    \chi\left(\omega, \Omega_c, \theta\right) = \sum_I \frac{|\mathcal{M}_{I,0}|^2}{\omega - (E_I \left(\Omega_c, \theta\right) - E_0 \left(\Omega_c, \theta\right)) + i\eta}
\end{equation}
where $\mathcal{M}_{I,0} = \bra{\Psi_I}\hat{P} \otimes \mathbb{I} \cdot \bm{e}\ket{\Psi_0}$ is the transition matrix element between the polaritonic ground state $\Psi_0$ and an excited state $\Psi_I$. $\mathbb{I}$ represents the identity operator for the photonic part.
$\eta$ is a small artificial broadening, $\omega$ is the energy of the probe field, and $E_I \left(\Omega_c, \theta\right)$ is the energy of the polaritonic state (eigenvalue of the QED Hamiltonian). $\bm{e}$ represents the polarization of the probe field, and may be chosen arbitrarily.
This function depends directly on $\omega$, and indirectly on $\Omega_c$ and $\theta$ (the eigenvalues and eigenvector change depending on these two parameters).
The transition matrix element $\mathcal{M}_{I,0}$ represents the matter response to a probe field.

For spatially unstructured cavities, where we use a single effective mode description at $\bm{\Bar{q} = 0}$, the transition operator $\Hat{P}$ can be written as:
\[\Hat{P} = \Tilde{A}_{0,\bm{0}} \sum_{ll', ij} \sum_{\bm{Q}, \nu, \lambda} \mathcal{I}_{ll',ij,\bm{Q,0}}^{\nu, \lambda} \hat{X}_{ll',ij,\bm{0}}^{\nu \dagger} + h.c.\]
where $\mathcal{I}_{ll',ij,\bm{Q,0}}^{\nu, \lambda}$ is defined in Eq. \ref{eq:i_coeff}. This quantity was used for Fig. 2a and Fig. 3 of the main text.

For spatially structured cavities, where we use a multi effective mode description of the cavity modes, the the transition operator $\Hat{P}$ can be written as:
\[\Hat{P} = \sum_{\lambda, \bm{\Bar{q}}} \Tilde{A}_{0,\bm{\Bar{q}}} \sum_{ll'} \sum_{i \in C, j \in V} \sum_{\bm{Q}, \nu} \left[ \mathcal{I}_{ll',ij,\bm{Q,\Bar{q}}}^{\nu, \lambda} \hat{X}_{ll',ij,\bm{\Bar{q}}}^{\nu \dagger} + \mathcal{B}_{ll',ij,\bm{Q,\Bar{q}}}^{\nu, \lambda} \hat{X}_{ll',ij,\bm{Q+\Bar{q}}}^{\nu \dagger} \hat{X}_{ll',ij,\bm{Q}}^{\nu} + h.c. \right] \]
where $\mathcal{B}_{ll',ij,\bm{Q,\Bar{q}}}^{\nu, \lambda}$ is defined in Eq. \ref{eq:b_coeff}. This quantity was used for Fig. 5 of the main text.

\subsection{Spectral function}
\label{subsec:spectral_function}
We define the spectral function $\mathcal{S}\left(\omega, \Omega_c, \theta, \bm{Q}\right)$ as:
\begin{equation}
    \label{eq:spectral_function}
    \mathcal{S}\left(\omega, \Omega_c, \theta, \bm{Q}\right) = \sum_I \frac{\bra{\Psi_I}\hat{S}_{\bm{Q}}^\dagger\ket{\Psi_0}\bra{\Psi_0}\hat{S}_{\bm{Q}}\ket{\Psi_I}}{\omega - (E_I \left(\Omega_c, \theta\right) - E_0 \left(\Omega_c, \theta\right)) + i\eta}
\end{equation}
where all quantities and indexes have the same definition as in Section \ref{subsec:linear_response}.
We use the spectral function to obtain the band structure of the polaritonic system in the Moiré BZ.
Its physical meaning varies according to the definition of the operator $\Hat{S}_{\bm{Q}}$. 
If $\Hat{S}_{\bm{Q}}^\dagger = \Hat{X}_{\bm{Q}}^\dagger \otimes \mathbb{I}$, where $\mathbb{I}$ is the identity operator for the photon part, then the spectral function represents the probability of creating an exciton at the \textit{k-point} $\bm{Q}$ (starting from the polaritonic ground state).
This operator is able to capture the interesting features of a Moiré system in the case of a spatially unstructured cavity (i.e. where the photon momentum is zero). This formulation was used for Fig. 2(b-d) and Fig. 4 of the main text.

\begin{figure}
    \centering
    \includegraphics[width=0.25\linewidth]{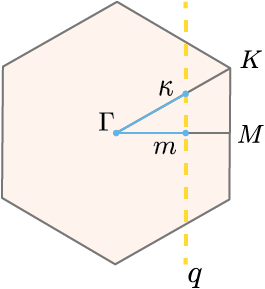}
    \caption{Representation of the $k$-points path used to plot the Spectral function in the Moiré Brillouin Zone. The yellow dashed line represents the momentum of the photonic mode $\bm{\bar{q}}$}
    \label{fig:k_path}
\end{figure}

When plotting the spectral function, we choose a specific path in the Moiré BZ. We use the standard $M-\Gamma-K$ path, which we shorten to $m-\Gamma-\kappa$, where $m, \kappa$ are taken at the value of the photon momentum along the aforementioned path. Such path is shown in Fig.~\ref{fig:k_path}

\section{Interaction Hamiltonian for classically driven cavities} \label{app:h_classical}
In this section we provide the full expression for the Interaction Hamiltonian (Eq. 6 of the main text) when we model the electromagnetic field inside a cavity as a time-dependent coherent state of light.
As in the main text, we will consider only two effective photonic modes in a spatially structured cavity, with momentum $\bm{q} = [q_x, 0], \bm{-q} = [-q_x, 0]$.
We write the photonic space as: 
\[
    \begin{cases}
        & \ket{\Tilde{\lambda}_{\bm{q}}, \Tilde{\lambda}_{\bm{-q}}, t} = e^{-\frac{i \left(\omega_{\bm{q}} + \omega_{\bm{-q}} \right) t}{2}} \ket{e^{-i \omega_{\bm{q}} t} \Tilde{\lambda}_{\bm{q}}, e^{-i \omega_{\bm{-q}} t} \Tilde{\lambda}_{\bm{-q}}} \\
        & \hat{a}_{\bm{q}} \ket{\Tilde{\lambda}_{\bm{q}}, \Tilde{\lambda}_{\bm{-q}}, t} = \Tilde{\lambda}_{\bm{q}} e^{-\frac{i \left(3\omega_{\bm{q}} + \omega_{\bm{-q}} \right) t}{2}} \ket{\Tilde{\lambda}_{\bm{q}}, e^{-i \omega_{\bm{-q}} t} \Tilde{\lambda}_{\bm{-q}}}
    \end{cases}
\]
where $\ket{\Tilde{\lambda}_{\bm{q}}, \Tilde{\lambda}_{\bm{-q}}, t} = \ket{\Tilde{\lambda}_{\bm{q}}, t} \otimes \ket{\Tilde{\lambda}_{\bm{-q}}, t}$ and $\ket{\Tilde{\lambda}_{\bm{q}}} = e^{-\frac{|\Tilde{\lambda}_{\bm{q}}|^2}{2}} \sum_{s=0}^{\infty} \frac{\Tilde{\lambda}^s}{\sqrt{s!}}\ket{s_{\bm{q}}}$.
Projecting the interaction Hamiltonian onto such a coherent state leads to:
\begin{equation}
    \label{eq:h_moire_eff_classical}
    \begin{aligned}
        & \bra{\Tilde{\lambda}_{\bm{q}}, \Tilde{\lambda}_{\bm{-q}}, t} \Hat{H}_{\text{int}} \ket{\Tilde{\lambda}_{\bm{q}}, \Tilde{\lambda}_{\bm{-q}}, t} = \sum_{ll'} \sum_{i \in C, j \in V} \sum_{\bm{Q}, \nu} \sum_{\lambda} \Biggr[ \mathcal{M}_{ll',ij,\bm{Q,q}}^{\nu} \hat{X}_{ll',ij,\bm{Q+q}}^{\nu \dagger} \hat{X}_{ll',ij,\bm{Q}}^{\nu} +  \\
        & \Tilde{A}_{0,\bm{q}} \left( \mathcal{B}_{ll',ij,\bm{Q,q}}^{\nu, \lambda} \hat{X}_{ll',ij,\bm{Q+q}}^{\nu \dagger} \hat{X}_{ll',ij,\bm{Q}}^{\nu} \Tilde{\lambda}_{\bm{q}}^* + \mathcal{B}_{ll',ij,\bm{Q,-q}}^{\nu, \lambda} \hat{X}_{ll',ij,\bm{Q-q}}^{\nu \dagger} \hat{X}_{ll',ij,\bm{Q}}^{\nu} \Tilde{\lambda}_{\bm{-q}}^* \right) e^{i \omega_{\bm{q}} t} + \\
        & \Tilde{A}_{0,\bm{q}} \left( \mathcal{B}_{ll',ij,\bm{Q,-q}}^{\nu, \lambda} \hat{X}_{ll',ij,\bm{Q-q}}^{\nu \dagger} \hat{X}_{ll',ij,\bm{Q}}^{\nu} \Tilde{\lambda}_{\bm{q}} + \mathcal{B}_{ll',ij,\bm{Q,q}}^{\nu, \lambda} \hat{X}_{ll',ij,\bm{Q+q}}^{\nu \dagger} \hat{X}_{ll',ij,\bm{Q}}^{\nu} \Tilde{\lambda}_{\bm{-q}} \right) e^{-i \omega_{\bm{q}} t} + \\
        & \Tilde{A}_{0,\bm{q}} \left( \mathcal{I}_{ll',ij,\bm{Q,q}}^{\nu, \lambda} \hat{X}_{ll',ij,\bm{q}}^{\nu \dagger} \Tilde{\lambda}_{\bm{q}}^* + \mathcal{I}_{ll',ij,\bm{Q,-q}}^{\nu, \lambda} \hat{X}_{ll',ij,\bm{-q}}^{\nu \dagger} \Tilde{\lambda}_{\bm{-q}}^* + h.c. \right) e^{i \omega_{\bm{q}} t} + \\
        & \Tilde{A}_{0,\bm{q}} \left( \mathcal{I}_{ll',ij,\bm{Q,-q}}^{\nu, \lambda} \hat{X}_{ll',ij,\bm{-q}}^{\nu \dagger} \Tilde{\lambda}_{\bm{q}} + \mathcal{I}_{ll',ij,\bm{Q,q}}^{\nu, \lambda} \hat{X}_{ll',ij,\bm{q}}^{\nu \dagger} \Tilde{\lambda}_{\bm{-q}} + h.c. \right) e^{-i \omega_{\bm{q}} t} \Biggr]
    \end{aligned}
\end{equation}

\subsection{High frequency limit} \label{subsec:high_frequency}
In this section, we discuss the high frequency limit for the classically driven Hamiltonian.
In the following, we consider a classically driven cavity with $\tilde{A}_0 = 0.02a.u.$ and two modes with momentum $\bm{q} = [0.009, 0], \bm{-q} = [-0.009, 0]$.
Since we are interested in studying the effect of the classical driving, we set the Moiré potential $\mathcal{M}_{ll',ij,\bm{Qq}}^{\nu} = 0$.

In the main text, we state that in the high frequency regime the external driving cannot modify the parabolic dispersion.
In fact, the effective Floquet Hamiltonian is given by the Van Vleck expansion~\cite{VanVleck}:
\[ H_{\text{eff}} = H^{(n=0)} + \frac{\left[H^{(n=-1)}, H^{(n=1)}\right]}{\omega} + \mathcal{O}\left(\frac{1}{\omega^2}\right) \]
where $n$ is the Floquet frequency index.
Clearly, if $\omega \rightarrow \infty$ one has that $H_{\text{eff}} = H^{(n=0)}$.
We show such progression in Fig. \ref{fig:classic_light_progression}, where we plot the spectral function for a for various values of the modes frequency.
At small values of $\omega$, the positive and negative frequencies replicas from the Floquet Hamiltonian mix with the excitonic dispersion, generating a rich spectrum.
As $\omega$ increases, the spectral function pictures a parabolic dispersion, meaning that the exciton is unperturbed.

\begin{figure}
    \centering
    \includegraphics[width=\linewidth]{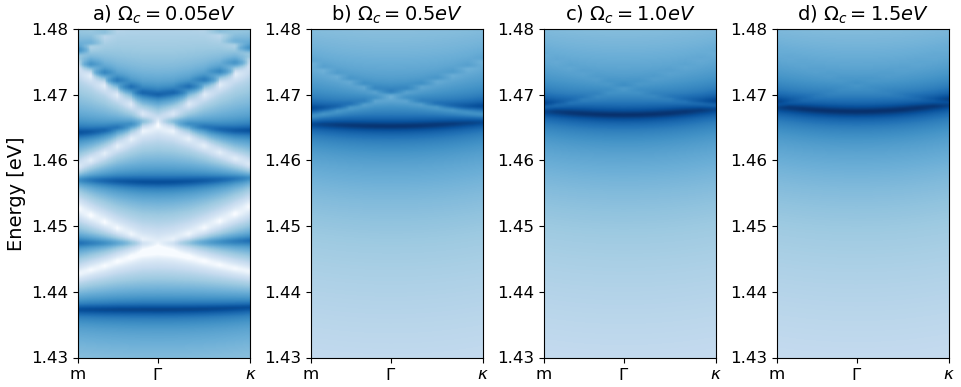}
    \caption{Spectral function for classically driven cavity with two modes, for various values of the modes' energy. In the high frequency regime, the effect of the effective Floquet Hamiltonian vanishes, as predicted by the Van Vleck expansion.}
    \label{fig:classic_light_progression}
\end{figure}

In the following, we provide the complete expression for the commutator $\left[H^{(n=-1)}, H^{(n=1)}\right]$. $H^{(n=1)}$ ($H^{(n=-1)}$) contains all terms from Eq. \ref{eq:h_moire_eff_classical} that are associated to $e^{i \omega_{\bm{q}} t}$ ($e^{-i \omega_{\bm{q}} t}$).
Note that since we are interested in studying the terms that originate the Moiré-like confinement, we will focus only on the conserving term of the bilinear coupling $\mathcal{B}_{ll',ij,\bm{Q},\bm{q}}^{\nu, \lambda}$. This is justified because the conserving term $\mathcal{I}_{ll',ij,\bm{Q},\bm{q}}^{\nu, \lambda}$ couples the ground state, and it signature on the spectrum is the Rabi splitting.
Hence:
\begin{align*}
    \left[H^{(n=-1)}, H^{(n=1)}\right] = & \sum_{ll'} \sum_{i \in C, j \in V} \sum_{\bm{Q}, \nu, \lambda} \Tilde{A}_{0, \bm{q}}^2 \Bigg[ \\
    & \Big( \mathcal{B}_{ll',ij,\bm{Q},\bm{-q}}^{\nu, \lambda} \mathcal{B}_{ll',ij,\bm{Q},\bm{q}}^{\nu, \lambda} |\Tilde{\lambda}_{\bm{q}}|^2 - \mathcal{B}_{ll',ij,\bm{Q},\bm{-q}}^{\nu, \lambda} \mathcal{B}_{ll',ij,\bm{Q},\bm{q}}^{\nu, \lambda} |\Tilde{\lambda}_{\bm{-q}}|^2 \Big) \hat{X}_{ll',ij,\bm{Q-q}}^{\nu \dagger} \hat{X}_{ll',ij,\bm{Q}}^{\nu} \hat{X}_{ll',ij,\bm{Q+q}}^{\nu \dagger} \hat{X}_{ll',ij,\bm{Q}}^{\nu} + \\
    & \Big( \mathcal{B}_{ll',ij,\bm{Q},\bm{-q}}^{\nu, \lambda} \mathcal{B}_{ll',ij,\bm{Q},\bm{q}}^{\nu, \lambda} |\Tilde{\lambda}_{\bm{-q}}|^2 - \mathcal{B}_{ll',ij,\bm{Q},\bm{-q}}^{\nu, \lambda} \mathcal{B}_{ll',ij,\bm{Q},\bm{q}}^{\nu, \lambda} |\Tilde{\lambda}_{\bm{q}}|^2 \Big) \hat{X}_{ll',ij,\bm{Q+q}}^{\nu \dagger} \hat{X}_{ll',ij,\bm{Q}}^{\nu} \hat{X}_{ll',ij,\bm{Q-q}}^{\nu \dagger} \hat{X}_{ll',ij,\bm{Q}}^{\nu} \Bigg]
\end{align*}
which is non-zero only when $|\Tilde{\lambda}_{\bm{q}}|^2 \neq |\Tilde{\lambda}_{\bm{-q}}|^2$.
Since for a coherent state of light one typically has that $|\Tilde{\lambda}_{\bm{q}}|^2 \approx |\Tilde{\lambda}_{\bm{-q}}|^2$, to observe a first order correction $\omega$ must be small.

\section{Hamiltonian down-folding} \label{app:downfolding}
This appendix discusses the downfolding of the QED Hamiltonian in a dressed photon space.
This approach allows us to define an effective Hamiltonian, composed of a single photon sector, that contains the action of the full QED Hamiltonian \cite{ferroelectric_gs}.
The first-order expansion of such an approximation for a system with $N$ modes can be written as~\cite{CavityGraphene}:
\begin{equation}
    \label{eq:h_downfolding_full}
    \begin{aligned}
        \hat{H}_{\text{eff}} & = \bra{0_1, \cdots, 0_N} \hat{H} \ket{0_1, \cdots, 0_N} - \\
        & \sum_\alpha^{N} \frac{1}{\omega_\alpha} \bigg[ \bra{0_1, \cdots, 0_N} \hat{H} \ket{0_1, \cdots, 1_\alpha, \cdots, 0_N} \cdot \bra{0_1, \cdots, 1_\alpha, \cdots, 0_N} \hat{H} \ket{0_1, \cdots, 0_N} \bigg] + \mathcal{O}\left(\frac{1}{\omega^2}\right)
    \end{aligned}
\end{equation}
where $\Hat{H}$ is the Hamiltonian one wants to approximate.

In the following we apply this expansion to the QED Hamiltonian (defined in the main text), for a system with two modes $\bm{q}$ and $\bm{-q}$. We will not focus on the free exciton Hamiltonian nor on the uncoupled photon Hamiltonian because we are interested on studying how the interacting terms are transformed by the down-folding.

Let us first consider the zeroth-order correction. The terms associated with the bilinear coupling have a photonic creation or annihilation operator, thus will not give any contribution. Hence:
\begin{equation}
    \label{eq:h_down_moire}
    \bra{0_{\bm{q}}, 0_{\bm{-q}}} \hat{H}_m \ket{0_{\bm{q}}, 0_{\bm{-q}}} = \sum_{ll'} \sum_{i \in C, j \in V} \sum_{\bm{Q}, \bm{q}, \nu} \mathcal{M}_{ll',ij,\bm{Q,q}}^{\nu}  \hat{X}_{ll',ij,\bm{Q+q}}^{\nu \dagger} \hat{X}_{ll',ij,\bm{Q}}^{\nu}
\end{equation}
where $\Hat{H}_{m}$ is the Moiré term in the QED Hamiltonian.

Let us now focus on the first-order correction. This time, the term associated with the Moiré potential will not give any contribution. Therefore, only the bilinear coupling terms will contribute (both the conserving, associated with the coefficient $\mathcal{B}$ and the non-conserving one, associated with the coefficient $\mathcal{I}$).

First, we consider the conserving term:
\[\Hat{H}_{bil, c} = \sum_{\lambda, \bm{\Bar{q}}} \Tilde{A}_{0,\bm{\Bar{q}}} \sum_{ll'} \sum_{i \in C, j \in V} \sum_{\bm{Q}, \nu} \mathcal{B}_{ll',ij,\bm{Q,\Bar{q}}}^{\nu, \lambda} \hat{X}_{ll',ij,\bm{Q+\Bar{q}}}^{\nu \dagger} \hat{X}_{ll',ij,\bm{Q}}^{\nu} \left(\hat{a}_{\bm{\Bar{q}},\lambda}^\dagger + \hat{a}_{-\bm{\Bar{q}}, \lambda}\right)\]
Since $\Hat{H}_{bil, c}$ contains a photonic creation and annihilation operator, the zeroth-order expansion yields
\newline $\bra{0_{\bm{q}}, 0_{\bm{-q}}} \Hat{H}_{bil, c} \ket{0_{\bm{q}}, 0_{\bm{-q}}} = 0$.
The first-order correction reads:
\begin{equation}
    \bra{0_{\bm{q}}, 0_{\bm{-q}}} \Hat{H}_{bil, c} \ket{1_{\bm{q}}, 0_{\bm{-q}}} = \Tilde{A}_{0,\bm{-q}} \sum_{ll'} \sum_{i \in C, j \in V} \sum_{\bm{Q}, \nu} \sum_{\lambda} \mathcal{B}_{ll',ij,\bm{Q,-q}}^{\nu, \lambda} \hat{X}_{ll',ij,\bm{Q-q}}^{\nu \dagger} \hat{X}_{ll',ij,\bm{Q}}^{\nu}
\end{equation}
\begin{equation}
    \bra{1_{\bm{q}}, 0_{\bm{-q}}} \Hat{H}_{bil, c} \ket{0_{\bm{q}}, 0_{\bm{-q}}} = \Tilde{A}_{0,\bm{q}} \sum_{ll'} \sum_{i \in C, j \in V} \sum_{\bm{Q}, \nu} \sum_{\lambda} \mathcal{B}_{ll',ij,\bm{Q,q}}^{\nu, \lambda} \hat{X}_{ll',ij,\bm{Q+q}}^{\nu \dagger} \hat{X}_{ll',ij,\bm{Q}}^{\nu}
\end{equation}
So:
\begin{equation}
    \begin{aligned}
        & \frac{\Hat{H}_{10,\bm{q}}^{bil, c} \Hat{H}_{10,\bm{q}}^{bil, c, \dagger}}{\omega_{\bm{q}}} = \frac{\bra{0_{\bm{q}}, 0_{\bm{-q}}} \hat{H}_{bil, c} \ket{1_{\bm{q}}, 0_{\bm{-q}}} \cdot \bra{1_{\bm{q}}, 0_{\bm{-q}}} \hat{H}_{bil, c} \ket{0_{\bm{q}}, 0_{\bm{-q}}}}{\omega_{\bm{q}}} = \\
        & \frac{\Tilde{A}_{0,\bm{q}} \Tilde{A}_{0,\bm{-q}}}{\omega_{\bm{q}}} \sum_{ll', l_1 l_1'} \sum_{ij, i_1 j_1} \sum_{\bm{Q}, \bm{Q}_1, \nu, \nu_1} \mathcal{B}_{ll',ij,\bm{Q,-q}}^{\nu, \lambda} \mathcal{B}_{l_1 l_1',i_1 j_1,\bm{Q_1,q}}^{\nu_1, \lambda}
        \hat{X}_{ll',ij,\bm{Q-q}}^{\nu \dagger} \hat{X}_{l_1 l_1',i_1 j_1,\bm{Q_1+q}}^{\nu_1 \dagger} \hat{X}_{ll',ij,\bm{Q}}^{\nu} \hat{X}_{l_1 l_1',i_1 j_1,\bm{Q_1}}^{\nu_1}
    \end{aligned}
\end{equation}
Equivalently, since $\omega_{\bm{q}} = \omega_{\bm{-q}}$ computing 
\[\frac{\Hat{H}_{10,\bm{-q}}^{bil, c} \Hat{H}_{10,\bm{-q}}^{bil, c, \dagger}}{\omega_{\bm{-q}}} = \frac{\bra{0_{\bm{q}}, 0_{\bm{-q}}} \Hat{H}_{bil, c} \ket{0_{\bm{q}}, 1_{\bm{-q}}} \cdot \bra{0_{\bm{q}}, 1_{\bm{-q}}} \Hat{H}_{bil, c} \ket{0_{\bm{q}}, 0_{\bm{-q}}}}{\omega_{\bm{-q}}}\]
leads to the same result. Hence, the full first-order correction for the conserving term reads:
\begin{equation}
    \label{eq:h_down_bil_c}
    \begin{aligned}
        & \frac{\Hat{H}_{10,\bm{q}}^{bil, c} \Hat{H}_{10,\bm{q}}^{bil, c, \dagger}}{\omega_{\bm{q}}} + \frac{\Hat{H}_{10,\bm{-q}}^{bil, c} \Hat{H}_{10,\bm{-q}}^{bil, c, \dagger}}{\omega_{\bm{-q}}} = \\
        & \frac{2 \Tilde{A}_{0,\bm{q}} \Tilde{A}_{0,\bm{-q}}}{\omega_{\bm{q}}} \sum_{ll', l_1 l_1'} \sum_{ij, i_1 j_1} \sum_{\bm{Q}, \bm{Q}_1, \nu, \nu_1} \mathcal{B}_{ll',ij,\bm{Q,-q}}^{\nu, \lambda} \mathcal{B}_{l_1 l_1',i_1 j_1,\bm{Q_1,q}}^{\nu_1, \lambda}
        \hat{X}_{ll',ij,\bm{Q-q}}^{\nu \dagger} \hat{X}_{l_1 l_1',i_1 j_1,\bm{Q_1+q}}^{\nu_1 \dagger} \hat{X}_{ll',ij,\bm{Q}}^{\nu} \hat{X}_{l_1 l_1',i_1 j_1,\bm{Q_1}}^{\nu_1}
    \end{aligned}
\end{equation}

Let us now expand the non-conserving term of the bi-linear coupling:
\[\Hat{H}_{bil, nc} = \sum_{\lambda, \bm{\Bar{q}}} \Tilde{A}_{0,\bm{\Bar{q}}} \sum_{ll'} \sum_{i \in C, j \in V} \sum_{\bm{Q}, \nu} \mathcal{I}_{ll',ij,\bm{Q,\Bar{q}}}^{\nu, \lambda} \hat{X}_{ll',ij,\bm{\Bar{q}}}^{\nu \dagger} \left(\hat{a}_{\bm{\Bar{q}},\lambda}^\dagger + \hat{a}_{-\bm{\Bar{q}}, \lambda}\right) + h.c.\]
Since $\Hat{H}_{bil, nc}$ contains a photonic creation and annihilation operator, the zeroth-order expansion yields 
\newline $\bra{0_{\bm{q}}, 0_{\bm{-q}}} \Hat{H}_{bil, nc} \ket{0_{\bm{q}}, 0_{\bm{-q}}} = 0$.
Conversely, the first-order correction reads:
\[\bra{0_{\bm{q}}, 0_{\bm{-q}}} \hat{H}_{bil, nc} \ket{1_{\bm{q}}, 0_{\bm{-q}}} = \Tilde{A}_{0,\bm{-q}} \sum_{ll'} \sum_{i \in C, j \in V} \sum_{\bm{Q}, \nu} \sum_{\lambda} \mathcal{I}_{ll',ij,\bm{Q,-q}}^{\nu, \lambda} \hat{X}_{ll',ij,\bm{-q}}^{\nu \dagger} + \Tilde{A}_{0,\bm{q}} \sum_{ll'} \sum_{i \in C, j \in V} \sum_{\bm{Q}, \nu} \sum_{\lambda} \mathcal{I}_{ll',ij,\bm{Q,q}}^{\nu, \lambda} \hat{X}_{ll',ij,\bm{q}}^{\nu}\]

\[\bra{1_{\bm{q}}, 0_{\bm{-q}}} \hat{H}_{bil, nc} \ket{0_{\bm{q}}, 0_{\bm{-q}}} = \Tilde{A}_{0,\bm{q}} \sum_{ll'} \sum_{i \in C, j \in V} \sum_{\bm{Q}, \nu} \sum_{\lambda} \mathcal{I}_{ll',ij,\bm{Q,q}}^{\nu, \lambda} \hat{X}_{ll',ij,\bm{q}}^{\nu \dagger} + \Tilde{A}_{0,\bm{-q}} \sum_{ll'} \sum_{i \in C, j \in V} \sum_{\bm{Q}, \nu} \sum_{\lambda} \mathcal{I}_{ll',ij,\bm{Q,-q}}^{\nu, \lambda} \hat{X}_{ll',ij,\bm{-q}}^{\nu}\]

So:
\begin{equation}
    \begin{aligned}
        & \frac{\Hat{H}_{10,\bm{q}}^{bil, nc} \Hat{H}_{10,\bm{q}}^{bil, nc, \dagger}}{\omega_{\bm{q}}} = \frac{\bra{0_{\bm{q}}, 0_{\bm{-q}}} \hat{H}_{bil, nc} \ket{1_{\bm{q}}, 0_{\bm{-q}}} \cdot \bra{1_{\bm{q}}, 0_{\bm{-q}}} \hat{H}_{bil, nc} \ket{0_{\bm{q}}, 0_{\bm{-q}}}}{\omega_{\bm{q}}} = \\
        & \frac{\Tilde{A}_{0,\bm{q}} \Tilde{A}_{0,\bm{-q}}}{\omega_{\bm{q}}} \sum_{ll', l_1 l_1'} \sum_{ij, i_1 j_1} \sum_{\bm{Q}, \bm{Q}_1, \nu, \nu_1} \mathcal{I}_{ll',ij,\bm{Q,q}}^{\nu, \lambda} \mathcal{I}_{l_1 l_1',i_1 j_1,\bm{Q_1,q}}^{\nu_1, \lambda, *} \hat{X}_{ll',ij,\bm{q}}^{\nu \dagger} \hat{X}_{l_1 l_1',i_1 j_1,\bm{q}}^{\nu_1} + \\
        & \frac{\Tilde{A}_{0,\bm{q}} \Tilde{A}_{0,\bm{-q}}}{\omega_{\bm{q}}} \sum_{ll', l_1 l_1'} \sum_{ij, i_1 j_1} \sum_{\bm{Q}, \bm{Q}_1, \nu, \nu_1} \mathcal{I}_{ll',ij,\bm{Q,-q}}^{\nu, \lambda, *} \mathcal{I}_{l_1 l_1',i_1 j_1,\bm{Q_1,-q}}^{\nu_1, \lambda} \hat{X}_{ll',ij,\bm{-q}}^{\nu} \hat{X}_{l_1 l_1',i_1 j_1,\bm{-q}}^{\nu_1 \dagger}
    \end{aligned}
\end{equation}
where we disregarded the terms involving a double creation or annihilation excitonic operators. This can be rewritten as:
\begin{equation}
    \begin{aligned}
        & \frac{\Hat{H}_{10,\bm{q}}^{bil, nc} \Hat{H}_{10,\bm{q}}^{bil, nc, \dagger}}{\omega_{\bm{q}}} = \frac{\bra{0_{\bm{q}}, 0_{\bm{-q}}} \hat{H}_{bil, nc} \ket{1_{\bm{q}}, 0_{\bm{-q}}} \cdot \bra{1_{\bm{q}}, 0_{\bm{-q}}} \hat{H}_{bil, nc} \ket{0_{\bm{q}}, 0_{\bm{-q}}}}{\omega_{\bm{q}}} = \\
        & \frac{\Tilde{A}_{0,\bm{q}} \Tilde{A}_{0,\bm{-q}}}{\omega_{\bm{q}}} \sum_{ll', l_1 l_1'} \sum_{ij, i_1 j_1} \sum_{\bm{q}, \bm{Q}, \bm{Q}_1, \nu, \nu_1} \mathcal{I}_{ll',ij,\bm{Q,q}}^{\nu, \lambda} \mathcal{I}_{l_1 l_1',i_1 j_1,\bm{Q_1,q}}^{\nu_1, \lambda, *} \hat{X}_{ll',ij,\bm{q}}^{\nu \dagger} \hat{X}_{l_1 l_1',i_1 j_1,\bm{q}}^{\nu_1}
    \end{aligned}
\end{equation}
Equivalently, since $\omega_{\bm{q}} = \omega_{\bm{-q}}$ computing 
\[\frac{\Hat{H}_{10,\bm{-q}}^{bil, nc} \Hat{H}_{10,\bm{-q}}^{bil, nc, \dagger}}{\omega_{\bm{-q}}} = \frac{\bra{0_{\bm{q}}, 0_{\bm{-q}}} \hat{H}_{bil, nc} \ket{0_{\bm{q}}, 1_{\bm{-q}}} \cdot \bra{0_{\bm{q}}, 1_{\bm{-q}}} \hat{H}_{bil, nc} \ket{0_{\bm{q}}, 0_{\bm{-q}}}}{\omega_{\bm{-q}}}\]
leads to the same result. Hence, the full first-order correction for the non-conserving term reads:
\begin{equation}
    \label{eq:h_down_bil_nc}
    \frac{\Hat{H}_{10,\bm{q}}^{bil, nc} \Hat{H}_{10,\bm{q}}^{bil, nc, \dagger}}{\omega_{\bm{q}}} + \frac{\Hat{H}_{10,\bm{-q}}^{bil, nc} \Hat{H}_{10,\bm{-q}}^{bil, nc, \dagger}}{\omega_{\bm{-q}}} = \frac{2 \Tilde{A}_{0,\bm{q}} \Tilde{A}_{0,\bm{-q}}}{\omega_{\bm{q}}} \sum_{ll', l_1 l_1'} \sum_{ij, i_1 j_1} \sum_{\bm{q}, \bm{Q}, \bm{Q}_1, \nu, \nu_1} \mathcal{I}_{ll',ij,\bm{Q,q}}^{\nu, \lambda} \mathcal{I}_{l_1 l_1',i_1 j_1,\bm{Q_1,q}}^{\nu_1, \lambda, *} \hat{X}_{ll',ij,\bm{q}}^{\nu \dagger} \hat{X}_{l_1 l_1',i_1 j_1,\bm{q}}^{\nu_1}
\end{equation}
Finally, the full down-folded Hamiltonian can be obtained by summing Eq. \ref{eq:h_down_moire}, \ref{eq:h_down_bil_c} and \ref{eq:h_down_bil_nc}.

\section{Methods and Computational details} \label{app:methods}
\subsection{Mott-Wannier model computational details}
Excitons in the Mott-Wannier model are formed thanks to the solution of Eq. \ref{eq:wannier}:
\[\frac{\hbar^2 k^2}{2 m_{ll'}^r} \psi_{ll'}^{\nu}\left(\bm{k}\right) - \sum_{\bm{q}} \mathcal{W}_{\bm{q}}^{ll'} \psi_{ll'}^{\nu}\left(\bm{k + q}\right) = \hat{E}_{ll'}^{\nu} \psi_{ll'}^{\nu}\left(\bm{k}\right)\]
where $\mathcal{W}_{\bm{q}}^{ll'}$ is the Coulomb potential defined in \ref{eq:h_start} and $\hat{E}_{ll'}^{\nu}$ is the bound energy.
To solve this equation, we model the Coulomb potential after Eq. 2 of the Supplementary Information of Ref.~\cite{2d_coulomb}. Table \ref{tab:wannier_numerical_vals} reports the numerical values used.
We build a $65 \times 65$ \textit{k-points} grid around $\Gamma$ and after solving equation we obtain the excitonic wavefunctions $\psi_{ll'}^\nu (\bm{k})$ and the bound energies $\hat{E}_{ll'}^{\nu}$.

\begin{table}[t]
    \centering
    \begin{tabular}{c|c|c|c|c|c}
        \hline
        Value & Sub1 & $MoSe_2$ & Interlayer & $WSe_2$ & Sub2 \\ 
        \hline
        Thickness [A] &  & 6.2926 & 6.22705 & 6.1615 &  \\
        Lattice param [A] &  & 3.32 &  & 3.319 &  \\
        Dielectric const & 4 & 16.5 & 1 & 15.1 & 4 \\
        Hole eff mass &  & 0.6 &  & 0.36 &  \\
        Elec eff mass &  & 0.5 &  & 0.29 &  \\
    \end{tabular}
    \caption{The numerical values (in atomic units) for the parameters required to compute the Coulomb potential.}
    \label{tab:wannier_numerical_vals}
\end{table}

\subsection{QED Hamiltonian approximations}
The QED Hamiltonian is reported in Eq. 3 of the main text.
We represent this Hamiltonian on the basis $\ket{\Psi_ {EX}} \otimes \ket{n}_0 \otimes \ket{n}_1 ...$, where $\ket{\Psi_ {EX}}$ is a Slater determinant representing an excitonic excitation or the many-body ground state and $\ket{n}_i$ represents a cavity mode. For each mode, we only consider the vacuum and one-photon state: $\{\ket{0}, \ket{1}\}$.
In the numerical simulations of this work, we make approximations to reduce the sums in the QED Hamiltonian.
In particular, we consider only the first excitonic state, thus $\nu = 1s$.
We also consider only one valence band and one conduction band, thus the indexes $i, j$ can be disregarded. 
All simulations presented in the paper are done for the intra-layer exciton of the MoSe$_2$ layer, thus also the indexes $l, l'$ can be dropped. Note that we tried to simulate the coupling to all excitons together, and we noticed that the physics of each is not affected by the presence of other excitons (at least to a first order approximation).
Despite not including the two layers explicitly, the system can still be considered a bi-layer because the effect of WSe$_2$ on MoSe$_2$ is included in the excitonic states $\psi(\bm{k})$ and corresponding eigenenergies and in the Moiré potential $\mathcal{M}_{\bm{Q,q}}$.
It is important to note that in principle one should not separate the different types of excitons as they can mix through the interaction with light. However, we verified that for the sake of what presented in this work, that mixing is not relevant.
Finally, we absorbed the diamagnetic term into the uncoupled photon Hamiltonian by performing a Bogoliubov transformation \cite{rokaj2022free, CavityGraphene}.
\newline Considering all assumptions above, the Hamiltonian we implemented in our code is the following:
\begin{equation}
    \label{eq:h_qed_simplified}
    \begin{aligned}
        & \hat{H}_{QED} = \sum_{\bm{\Bar{q}},\lambda} \omega_{\bm{\Bar{q}}} \left(\hat{a}_{\bm{\Bar{q}}, \lambda}^\dagger \hat{a}_{\bm{\Bar{q}}, \lambda} + \frac{1}{2}\right) + \sum_{\bm{Q}} \mathcal{E}_{\bm{Q}} \hat{X}_{\bm{Q}}^\dagger \hat{X}_{\bm{Q}} + \sum_{\bm{Q,q}} \mathcal{M}_{\bm{q}} \hat{X}_{\bm{Q+q}}^\dagger \hat{X}_{\bm{Q}} + \\
        & \sum_{\lambda, \bm{\Bar{q}}} \Tilde{A}_{0,\bm{\Bar{q}}} \sum_{\bm{Q}} \mathcal{B}_{\bm{Q,\Bar{q}}}^{\lambda} \hat{X}_{\bm{Q+\Bar{q}}}^{\dagger} \hat{X}_{\bm{Q}} \left(\hat{a}_{\bm{\Bar{q}},\lambda}^\dagger + \hat{a}_{-\bm{\Bar{q}}, \lambda}\right) + \sum_{\lambda, \bm{\Bar{q}}} \Tilde{A}_{0,\bm{\Bar{q}}} \sum_{\bm{Q}} \mathcal{I}_{\bm{Q,\Bar{q}}}^{\lambda} \hat{X}_{\bm{Q}}^{\dagger} \left(\hat{a}_{\bm{\Bar{q}},\lambda}^\dagger + \hat{a}_{-\bm{\Bar{q}}, \lambda}\right) + h.c.
    \end{aligned}
\end{equation}
We solve this Hamiltonian by building its matrix representation on the aforementioned basis and performing an exact diagonalization, which gives us access to the polaritonic eigenvalues and eigenstates.

\subsection{Transition matrix elements}
\label{subsec:transition_me}
The transition matrix elements are given by computing the quantity 
\[\bra{\Psi_{EX, \bm{Q}}, n_{\Bar{\bm{q}}}} \Hat{H}_{bil} \ket{\Psi_{EX, \bm{q'}}, m_{\Bar{\bm{q}}}}\]
where $\ket{n_{\bm{\Bar{q}}}}, \ket{m_{\bm{\Bar{q}}}}$ represent the initial and final photonic state for the mode $\bm{\Bar{q}}$, and $\Hat{H}_{bil}$ is:
\[\Hat{H}_{bil} = \sum_{\lambda, \bm{\Bar{q}}} \Tilde{A}_{0,\bm{\Bar{q}}} \sum_{\bm{Q}} \mathcal{B}_{\bm{Q,\Bar{q}}}^{\lambda} \hat{X}_{\bm{Q+\Bar{q}}}^{\dagger} \hat{X}_{\bm{Q}} \left(\hat{a}_{\bm{\Bar{q}},\lambda}^\dagger + \hat{a}_{-\bm{\Bar{q}}, \lambda}\right) + \sum_{\lambda, \bm{\Bar{q}}} \Tilde{A}_{0,\bm{\Bar{q}}} \sum_{\bm{Q}} \mathcal{I}_{\bm{Q,\Bar{q}}}^{\lambda} \hat{X}_{\bm{Q}}^{\dagger} \left(\hat{a}_{\bm{\Bar{q}},\lambda}^\dagger + \hat{a}_{-\bm{\Bar{q}}, \lambda}\right) + h.c.\]
As for the photonic part, it is immediate to see that the matrix element is non-zero only if $\ket{n_{\bm{\Bar{q}}}} = \ket{m_{\bm{\Bar{q}}}} \pm 1$.
As for the matter part, we shall distinguish between $\bm{\Bar{q}}$ equal to zero (only vertical transitions are allowed) or finite (the photon transfers its momentum to the matter). 
In the former case, which throughout the paper is referred to as a spatially unstructured cavity, one has that only 
\[\mathcal{I}_{\bm{Q,\Bar{q}}}^{\lambda} = \sum_{\bm{k}} p_{\bm{k}+\bm{\Bar{q}}, \bm{k}}^\lambda \psi \left[\alpha\left(\bm{k}\right) + \beta\left(\bm{k-Q}\right)\right]\]
contributes to the coupling. In TMDs, excitons are very localized in some points of the Brillouin zone, thus one can approximate that the valence-conduction momentum element $p_{\bm{k}+\bm{\Bar{q}}, \bm{k}}$ is constant with $\bm{k}$. In particular, we choose its value at \textit{K} as reference.
Thus, we compute $\mathcal{I}_{\bm{Q,\Bar{q}}}^{\lambda} = p_{K+\bm{\Bar{q}}, K}^\lambda \sum_{\bm{k}} \psi \left[\alpha\left(\bm{k}\right) + \beta\left(\bm{k-Q}\right)\right]$.
\newline In contrast, if $\bm{\Bar{q}}$ is not zero, which throughout the paper is referred to as a spatially structured cavity, one should also account for
\[\mathcal{B}_{\bm{Q,\Bar{q}}}^{\lambda} = \sum_{\bm{k}} \left[p_{\bm{k+\Bar{q},\bm{k}}}^{cc, \lambda} \Psi\left(\bm{k-Q}, \bm{k+\Bar{q}, \bm{k-Q}, \bm{k}}\right) - p_{\bm{k+\Bar{q},\bm{k}}}^{vv, \lambda} \Psi \left(\bm{k}, \bm{k-Q}, \bm{k+\Bar{q}}, \bm{k-Q}\right)\right]\]
By assuming that the valence and conduction band are parabolic, we can simplify the expression above as $p_{\bm{k+\Bar{q},\bm{k}}}^{cc, \lambda} = -p_{\bm{k+\Bar{q},\bm{k}}}^{vv, \lambda} = \frac{\norm{\bm{\Bar{q}}}}{m_{\textit{eff}}}$ \cite{cavity_control}, where $m_{\textit{eff}}$ is the excitonic effective mass. Hence, 
\[\mathcal{B}_{\bm{Q,\Bar{q}}} = \frac{\norm{\bm{\Bar{q}}}}{m_{\textit{eff}}} \sum_{\bm{k}} \left[ \Psi\left(\bm{k-Q}, \bm{k+\Bar{q}, \bm{k-Q}, \bm{k}}\right) + \Psi \left(\bm{k}, \bm{k-Q}, \bm{k+\Bar{q}}, \bm{k-Q}\right) \right]\]
\end{widetext}

\end{document}